\documentclass[11pt, letter]{article}
\pdfoutput=1

\usepackage{jheppub}

\usepackage[utf8]{inputenc}

\usepackage{mathtools}
\usepackage{dsfont}
\usepackage{mathdots}

\usepackage{tikz}
\usetikzlibrary{arrows,shapes}
\usetikzlibrary{trees}
\usetikzlibrary{patterns}
\usetikzlibrary{matrix,arrows} 				% 
\usetikzlibrary{positioning}				% For "above of=" commands
\usetikzlibrary{calc,through}				% For coordinates
\usetikzlibrary{decorations.pathreplacing}  % For curly braces
\usepackage{pgffor}							% For repeating patterns

\usetikzlibrary{decorations.pathmorphing}	% For Feynman Diagrams
\usetikzlibrary{decorations.markings}

\def\be{\begin{equation}}
\def\ee{\end{equation}}
\def\bea{\begin{eqnarray}}
\def\eea{\end{eqnarray}}
\def\ba{\begin{array}}
\def\ea{\end{array}}
\def\bd{\begin{displaymath}}
\def\ed{\end{displaymath}}

\def\>{\rangle} %right angle
\def\<{\langle} %left angle
\def\Dsl{D \hskip-.6em \raise1pt\hbox{$ / $ } }
\def\to{\rightarrow}

\title{Thermodynamics of Near BPS Black Holes in AdS$_4$ and AdS$_7$ }

\author{Finn Larsen}
\author{and Shruti Paranjape}

\affiliation{Leinweber Center for Theoretical Physics,\\ 
Department of Physics, University of Michigan,\\
450 Church St, Ann Arbor, MI 48109, USA}

\emailAdd{larsenf@umich.edu}
\emailAdd{shrpar@umich.edu}

\abstract{We develop the thermodynamics of black holes in AdS$_4$ and AdS$_7$
near their BPS limit. In each setting we study the two distinct deformations orthogonal to the BPS surface as well as their nontrivial interplay with each other and with BPS properties. Our results illuminate recent microscopic calculations of the BPS entropy. We show that these microscopic computations can be leveraged to also describe the near BPS regime, by generalizing the boundary conditions imposed on states.

\vspace{2cm} 
\noindent {\tt LCTP-20-24}}


\begin{document}  
%%%%%%%%%%%%%%%%%%%%%%%%%%%%%%%%%%%%%%%

\maketitle

\newpage

%%%%%%%%%%%%%%%%%%%%%%%%%%%%%%
\section{Introduction}
\label{sec:Introduction}
Physicists have recently made significant advances towards a microscopic understanding of black hole entropy in AdS spacetimes \cite{Choi:2018hmj,Cabo-Bizet:2018ehj,Benini:2018ywd}.
Nearly all progress has relied heavily on supersymmetry, such as
using the supersymmetric index to count states or supersymmetric localization to compute the effective action. 
These methods are powerful and quite rigorous, but they also have obvious limitations. For example,
some physical black hole properties change discontinuously in the strict supersymmetric limit \cite{Preskill:1991tb}.

In this paper we study {\it nearly} supersymmetric black holes in AdS. Such black holes are important because they have many physical properties in common with 
generic black holes, yet they inherit some of the technical advantages held by their strictly supersymmetric relatives. 
More precisely, our specific goal is to develop properties of nearBPS black holes in AdS$_4$ and AdS$_7$ and compare with analogous results previously established in AdS$_5$ \cite{Larsen:2019oll}. Some details differ between these settings, of course, but several aspects are so similar that they may be described by the same effective theory. This agrees nicely 
with the understanding of universality emerging in the context of nearAdS$_2$/nearCFT$_1$ correspondence, another research direction with rapid progress over the last few years \cite{Almheiri:2014cka,MSZ,Kitaev:2017awl}.

\subsection{SUSY or Not?}
Given the central importance of supersymmetry in nearly all current work in the area we now address, 
before getting to further details, how any progress can be made at all. To do so, recall the microstate counting of 
asymptotically flat black holes which has a long history and is understood incomparably better. 
Many precision agreements were established, not just at the leading order but also for higher derivative corrections, quantum corrections, and far beyond \cite{Castro:2008ne,Dabholkar:2010rm,Sen:2014aja}. 
Moreover, in most cases it has been understood {\it why} these agreements hold with the precision they do. The reasons vary according to the setting and, although they
often involve supersymmetry, that is not always the case. In particular, it has proven fruitful to study black holes that are solutions to theories with supersymmetry 
without themselves preserving any supersymmetry.

Specifically, experience with asymptotically flat black holes suggests that small deformations away from the BPS limit are under 
good control. One avenue is revealed geometrically by the near horizon AdS$_2$ being enlarged to an AdS$_3$. In this situation a combination of 
anomaly arguments and modular invariance ensures agreements at leading and subleading order, even when supersymmetry is broken \cite{Maldacena:1997de,Kraus:2005vz}. This success is not a feature of AdS$_3$ alone, the entropy
of extremal but non-BPS black holes can be accounted for correctly even at four-derivative level, as shown by application of the entropy extremization formalism 
in AdS$_2$ \cite{Banerjee:2016qvj}. Such successes for asymptotically flat black holes motivate studying nearly supersymmetric black holes in AdS. 

\subsection{nAdS$_2/$CFT$_1$ Correspondence}
As we mentioned tangentially already, a somewhat complementary motivation for studying near BPS black holes is presented by recent progress on their holographic description 
through the nearAdS$_2$/nearCFT$_1$ correspondence \cite{MSZ,Almheiri:2016fws}. 
A central aspect of this duality is a nontrivial symmetry breaking pattern 
which coincides between the two sides. It is realized in 
melonic quantum theories such as the SYK model and its avatars \cite{Sachdev:1992fk,Kitaev,Maldacena:2016hyu}, novel settings that have justifiably generated much interest. Importantly, the symmetry breaking pattern is also realized in gravitational theories such as the Jackiw-Teitelboim model and its relatives \cite{Almheiri:2014cka,MSZ,Kitaev:2017awl}. This sets the scene for a holographic duality. 

However, nearAdS$_2$/nearCFT$_1$ correspondence is not a straightforward equivalence, it is an IR duality where the effective theory in the bulk and on the boundary are dual to one another only at large distances. The significance of the near BPS AdS black holes we study is that they offer a UV completion of the description on both sides of the duality. Thus, for each of the black holes, there is a specific dual theory that is well-defined in the UV. 

The most precise studies of black holes focus on supersymmetric ground states and their ability to describe the entropy of BPS black holes microscopically. This is very interesting, of course, but BPS states are relatively inert ground states, unsuitable as proxies for physical black holes. It is 
therefore important to explore the low lying excited states as well \cite{Larsen:2019oll}. That is one of our motivations. 

The low energy effective theory is usefully summarized as a theory of Schwarzian type, characterized by one or more dimensionful coupling constants. These coefficients are response parameters such as the specific heat of the black hole. They are arbitrary inputs from the effective field theory point of view. However, in the context of a UV complete theory 
the response parameters can be determined from microscopic principles \cite{Moitra:2018jqs,Castro:2018ffi,Moitra:2019bub,Castro:2019crn}. One goal of this article is to do so explicitly. 

\subsection{Black Holes in AdS}
The bosonic symmetry groups of AdS$_4\times S^7$, AdS$_5\times S^5$, and AdS$_7\times S^4$ all have rank $6$. Therefore, in each case the BPS condition derived from supersymmetry becomes a linear relation between $6$ conserved charges that arise as eigenvalues of their respective Cartan generators. In other words, it expresses the mass $M$ as a sum of $5$ conserved charges as:
\begin{equation}
\label{eqn:BPSmassintro}
M = \sum_I Q_I + \sum_i J_i~,   
\end{equation}
in appropriate conventions.
Here the range of the indices $I$ and $i$ depend on the spacetime dimension but in all cases they enumerate a total of $5$ values.  

However, it turns out that regular black holes in AdS only exist if, in addition, a certain constraint is imposed on the $5$ charges $(Q_I,J_i)$. For example, BPS black holes in gauged ${\cal N}=8$ supergravity in five dimensions are characterized by $5$ charges that satisfy the constraint 
\begin{equation}
\small
\label{eqn:constraintintro}
h = Q_1 Q_2 Q_3  + \frac{1}{2}N^2 J_1 J_2
- \left( \frac{1}{2}N^2 + Q_1 + Q_2 + Q_3\right)\left(Q_1 Q_2 +
Q_2 Q_3+ Q_3 Q_1 - \frac{1}{2}N^2(J_1+J_2) \right) = 0 ~,
\end{equation}
\normalsize
where $N^2$ refers to the dual ${\cal N}=4$ SYM theory with $SU(N)$ gauge group. This type of constraint may at first appear novel and special to AdS but actually it is not, it is just more complicated in AdS than in the more familiar settings with asymptotically flat spacetime. For example, within the family of Kerr-Newman black holes in four-dimensional Einstein-Maxwell theory (supersymmetrized as minimal ungauged ${\cal N}=2$ supergravity), the BPS black holes have mass given in terms of charge as $M = Q$ {\it and} they have angular momentum $J=0$. There are certainly regular extremal Kerr-Newman black holes that rotate, including the extremal Kerr black hole that is neutral under the gauge field. Alas, supersymmetry demands more than vanishing temperature, it imposes the constraint that the angular momentum $J=0$. 

The recognition of the constraint plays an important role in this article. We study thermodynamics of near BPS black holes, i.e. parameter values that differ slightly from those of a reference BPS black hole. Importantly, there is a {\it two}-dimensional space of deformations. The most obvious is to increase the black hole mass beyond its BPS value while keeping charges fixed. This is equivalent to increasing temperature to $T>0$. 
The alternative is to maintain extremality ($T=0$) but modify charges so they violate the constraint. 

The two directions away from the BPS surface both have a preferred sign: stability imposes not only $T\geq 0$ but also $h\geq 0$. Thus the parameter space is a quadrant of the plane with the BPS point at the origin. Moreover, the interplay between the two directions is quite nontrivial. It is characterized by $3$ response functions, each of which depend on all the conserved charges (subject to the constraint). A successful microscopic theory of the near BPS black holes must account for all these nontrivial functions. 

\subsection{Microscopic Description of near BPS AdS Black Holes}
In this article we do not attempt to derive a microscopic theory of the AdS black holes we study {\it ab initio} but we explain how recent progress on the strict BPS limit can be leveraged towards that goal. 

The understanding of the AdS black hole entropy from microscopic principles has admittedly proven quite subtle even in the BPS limit. However, even though significant technical questions remain, the outline is now generally agreed upon. Accordingly, we take it as a given that the entropy of BPS black holes in AdS can be interpreted in terms of a dual field theory. We think of the required estimate of the asymptotic density of states as a two step-process: enumeration of all states in a ``large" Hilbert space, followed by identifying physical states as
those satisfying the constraint \eqref{eqn:constraintintro}, or one of its analogues in other dimensions, 

Our strategy for addressing near BPS black holes is to take the ``large" Hilbert space established in the course of investigating BPS black holes as a given starting point. We then identify physical states by imposing a constraint that has been relaxed to accommodate a departure from the BPS limit. The less demanding constraint permits more physical states and this allows computation of the excess entropy enjoyed by near BPS black holes. We find that the entropy computed from this microscopic reasoning agrees with the one found from gravitational thermodynamics. 

Our prescription describing near BPS thermodynamics is intuitive and physically reasonable but it goes against conventional wisdom on what quantities can be reliably computed. It suggests that agreements are justified not by preserved supersymmetry alone but also by broken supersymmetry and/or anomalies. 
We stress again that we do not claim fully principled comparisons between nonsupersymmetric black holes in AdS and their holographically dual boundary theory. It would in fact be premature to expect such results since the BPS agreements themselves remain beset by questions. 
However, the 
agreements we report are quite impressive as they involve many parameters and apply in each of the dimensions we develop. This persuades us that they are accurate and we expect they will ultimately acquire a solid justification. 

\subsection{Organization of This Article}

This paper is divided into two parts: section \ref{sec:AdS4} on 
AdS$_4$ black holes and section \ref{sec:AdS7} on AdS$_7$ black holes. We have purposely written these sections on AdS$_4$ and AdS$_7$ so they are largely independent and can be studied in any order. 

Within each section, we first discuss black hole thermodynamics generally and then consider the nature of the BPS limit. This sets up the development of several distinct near BPS regions and their interplay, all from the gravitational point of view. Our discussion of the microscopic description is collected in subsection \ref{sec:AdS4mic} for AdS$_4$ and subsections 
\ref{sec:BPSmic}-\ref{subsec:mic7}
for AdS$_7$.

%%%%%%%%%%%%%%%%%%%%%%%%%%%%%%

\section{The Kerr-Newman-AdS\texorpdfstring{\textsubscript{4}}{4} Black Hole}
\label{sec:AdS4}
In this section we study the thermodyamics of Kerr-Newman AdS$_4$ black holes. We discuss the constraint on charges or potentials that is required for supersymmetry
and consider the nearBPS black holes that have small temperature and/or fail to satisfy the constraint. We show that the
partition function that accounts for BPS black hole entropy microscopically also describes the nearBPS regime.

%%%%%t
\subsection{The Kerr-Newman AdS\texorpdfstring{$_4$}{4} Black Hole}
The $6$ quantum numbers of the maximally supersymmetric black holes in AdS$_4$
are the mass $M$, the angular momentum $J$, and four $R$-charges $Q_I$ ($I=1,2,3,4$) that correspond to the Cartan generators of the $S^7$ isometry group $SO(8)$. We specialize to the Kerr-Newman AdS$_4$ black holes where the four R-charges are identical so the solution depends on just three 
quantum numbers: $M$, $J$, and $Q$. The only other parameters that enter are the asymptotically AdS$_4$ radius $\ell_4$ (related to the coupling $g=\ell_4^{-1}$ 
of gauged supergravity) and the gravitational coupling $G_4$. 

The explicit solution (first presented in \cite{Chong:2004dy}) is fairly elaborate, as expected for a rotating black hole, so we will not present the geometry and its associated matter here. 
The only feature that is needed in our study is the radial function 
\begin{equation}
\label{eq:horeqn}
	\Delta_r (r) = (r^2 + a^2) (1 + g^2 r^2) - 2mr + q^2~,
\end{equation}
that appears prominently in the metric. 
The event horizon of the black hole is located at the coordinate $r=r_+$ that is the largest real root of the quartic equation $\Delta_r(r)=0$. 
The parameters $(m, a, q)$ in $\Delta_r$ are related to the physical variables $(M, J, Q)$ of the 
black hole as \cite{Caldarelli:1999xj,Gibbons:2004ai,Papadimitriou:2005ii}
\begin{eqnarray}
\label{eq:QJM}
	M & = &  \frac{1}{G_4} \frac{m}{\Xi^2}~,\cr 
	J & = &  \frac{1}{G_4} \frac{ma}{\Xi^2}~,\cr
Q & = &  \frac{1}{2G_4}  \frac{q}{\Xi}~,
\end{eqnarray}
where $\Xi = 1 -  a^2g^2$. The parameters $m, q$ are positive while $0\leq ag<1$. The charges are normalized so $2Q\ell_4$ and $J$ are integral (for bosons) or half-integral (for fermions). 

In black hole thermodynamics a central role is played by the 
potentials that are conjugate to the three quantum numbers $(M, J, Q)$. They are: 
\begin{eqnarray}
\label{eq:TOP}
T & = & \frac{r_+}{4\pi ( r^2_+ + a^2)} \left( 1 + \frac{a^2 + 3r^2_+}{\ell^2_4} - \frac{a^2 + q^2}{r^2_+}\right)~,\cr
\Omega & = & \frac{a}{r_+^2 + a^2} \left(1 + g^2 r^2_+\right)~,\cr
\Phi & = & \frac{2qr_+}{r^2_+ + a^2}~. 
\end{eqnarray}
As usual, the size of the quantum configuration space is encoded in the black hole entropy
\begin{equation}
S =  \frac{\pi}{G_4} \frac{r^2_+ + a^2}{\Xi}~. 
\label{eq:ads4ent}
\end{equation}
We also record the on-shell Euclidean action of the black hole
\begin{equation}
	I = \frac{1}{2G_4} \frac{g^2}{\Xi T}\left( m\ell^2_4 - r_+ (r^2_+ + a^2) - \frac{q^2\ell^2_4 r_+}{r^2_+ + a^2} \right)  ~.
	\label{eq:onact}
\end{equation}
It satisfies the quantum statistical relation
$$
G \equiv T I = M - TS - \Omega J - \Phi Q  ~,
$$
as it should.

%%%%%%%%%%%%%%

\subsection{The BPS Bound}

The supersymmetry algebra realized by the theory demands that the black hole mass satisfies the BPS 
bound\footnote{In comparison with the schematic formula \eqref{eqn:BPSmassintro} each of the four Cartan generators are
taken as $\frac{1}{2}Q_I=Q_{\rm here}$. Also, here mass $M$ and charge $Q$ have dimension of inverse length while $J$ is a pure number.}
\begin{equation}
	\label{eqn:BPSphys}
	M \geq M^* = 2Q^* + g J^* ~,
\end{equation}
with the inequality saturated (i.e. satisfied as an equality) precisely when the black hole preserves a fraction of the supersymmetry. The * designates that variables refer to  
the BPS black holes. 
In this subsection we take the view that the BPS bound is a hypothesis that we seek to validate through explicit computation, by showing that it is satisfied by all the aforementioned black hole solutions. 

As a first step we rewrite the BPS bound (\ref{eqn:BPSphys}) using our parametric formulae \eqref{eq:QJM}
\begin{eqnarray}
	M - (2Q + gJ)  =  \frac{1}{G_4} \frac{1-ag}{\Xi^2}\left( m - q(1+ag)\right) \geq 0~.
\end{eqnarray}
Thus the inequality on physical variables is equivalent to the bound on parameters
\begin{equation}
	\label{eqn:BPSparam}
	m \geq (1+ag)q ~,
\end{equation}
with BPS saturation corresponding to equality in both cases. 

To make further progress we differentiate the BPS equality $M^*= 2Q^* +gJ^*$ and find the potentials $\Phi^*=2$, $\Omega^*=g$ for BPS configurations. 
We can invert the parametric formulae \eqref{eq:TOP} for these potentials to find the corresponding BPS relation between parameters 
\begin{eqnarray}
	g q^* &=& \sqrt{ag} \left( 1 + ag \right)~,
\end{eqnarray}
and also find the coordinate position of the horizon
\begin{eqnarray}
	g r^* &=& \sqrt{ag} ~.
\end{eqnarray}
It is a consistency check that the temperature $T=0$ for these values of parameters, as expected in the BPS limit. 

It follows from the facts established so far that the radial function $\Delta_r(r)$ \eqref{eq:horeqn} must vanish at $r=r^*$ for black hole parameters such that $q=q^*$ and
$m=m^*=(1+ag)q^*$. We can make this feature manifest by rewriting the general formula for the radial function as: 
\begin{eqnarray}
  	\Delta_r &=&  - 2\big[ m - q(1+ ag ) \big] r + (r-r^{*})^2\left(  1 + 6ag  + a^2 g^2   \right) - 
	2(q-q^*)(r-r^*)( 1 + a g  ) \cr
	&&~~~~+ (q-q^*)^{2} + 4\sqrt{ag}(r-r^{*})^3g + (r-r^{*})^4 g^2 ~.
\end{eqnarray}
The location of the horizon  $r=r_+$ is the largest solution to $\Delta_r(r)=0$ so,
for any value of black hole parameters, it satisfies the {\it exact} equation
\begin{align}
	2\big[ m - &q(1+a g )\big] r_+ \nonumber\\
	=&  \frac{1}{1 + 6 ag  + a^2 g^2} 
	\left[ (r_+-r^{*}) \left(  1 + 6 a g  + a^2 g^2 \right) - (q-q^*) ( 1 + a g )   + 
	2\sqrt{ag} g(r_+-r^{*})^2\right]^2  \nonumber\\
	&+  \frac{1}{1 + 6 a g +a^2 g^2 }  
	\left[ 2 \sqrt{ag}(q-q^*) + ( 1 + a g ) g  (r_+-r^{*})^2\right]^{2}~.
	\label{eqn:mexact}
\end{align}
The terms on the right hand side are manifestly positive so we conclude that the black hole parameters must satisfy  
$m - q(1+ag)\geq 0$. This agrees with the parametric bound \eqref{eqn:BPSparam}
so we have established by explicit computation that the physical BPS bound $M\geq M^*$ is satisfied for all the black holes solutions, as we wanted to show. 

This result was expected from supersymmetry of the theory. However, there is a less obvious corollary of the computation. 
The identity \eqref{eqn:mexact} shows that the BPS bound is saturated if and only if {\it both} of the square brackets on its right hand side vanish. 
This is clearly the case for ``the" BPS black holes with $q=q^*$ and $r_+=r^*$ that we have already identified but it is not difficult to check that this is the unique solution. 
In other words, we have shown by explicit computation that the BPS bound on the black hole mass \eqref{eqn:BPSphys} 
is saturated if and only if $q=q^*$ {\it and} $r_+=r^*$. We will see below that these conditions on parameters correspond to vanishing temperature $T=0$ {\it and} an additional constraint on the physical potentials
$\Phi-\Omega\ell_4 = 1$ or on the charges $Q$, $J$. 

%%%%%%%%%%%%%%%%%%%%%%%%%
\subsection{Formulae for BPS Black Holes}

We will discuss general black holes with frequent reference to the BPS limit. 
Therefore, we collect formulae for this  special case in this short subsection. We label the one-parameter family of BPS black holes by $ag$ and express the other dimensionless parameters as
\begin{eqnarray}
	g q^* &=& \sqrt{ag} \left( 1 + ag \right)~,\cr
	g r^* &=& \sqrt{ag}~,\cr
	g m^*  & = &  \sqrt{ag} \left( 1 + ag \right)^2 ~.
	\label{eq:qrmBPS}
\end{eqnarray}
Inserting these values into into \eqref{eq:QJM} we find the electric charge
\begin{eqnarray}
\label{eq:QJM-BPS}
	Q^* \ell_4 & = &  \frac{\ell^2_4}{2G_4}  \frac{\sqrt{ag} }{1 - ag} ~.
\end{eqnarray}
This formula can be inverted as
\begin{equation}
	ag = \left[ \frac{\ell_4}{4G_4 Q^*}\left( \sqrt{ 1  + \frac{16G^2_4Q^{*2}}{\ell^2_4 }}  - 1 \right)\right]^2~.
	\label{eq:agBPS}
\end{equation}
We can use this equation to eliminate the parameter $ag$ in favor of the charge when considering any physical variable of a BPS black hole. As an important example, after inserting the BPS parameters \eqref{eq:qrmBPS} into \eqref{eq:QJM} for the angular momentum, we can eliminate $ag$ and find
\begin{equation}
 J^* = Q^*\ell_4\left[ \sqrt{ 1  + \frac{16G^2_4Q^{*2}}{\ell^2_4 }}  - 1\right]~.
 \label{eq:Jconstraint}
\end{equation}
This relation between physical conserved charges is the {\it constraint} that must be satisfied for all BPS black holes. 

We can similarly find the BPS black hole entropy by substituting the parametric formulae \eqref{eq:qrmBPS} in the general equation for the entropy \eqref{eq:ads4ent} and then eliminate $ag$ using \eqref{eq:agBPS}. However, because of the constraint \eqref{eq:Jconstraint} the dependence 
of the BPS entropy on the conserved charges $Q^*, J^*$ is not unique, it can take many different forms.  
Our ``preferred" formula will be to eliminate the angular momentum entirely and express the black hole entropy it in terms of charges alone
\begin{eqnarray}
	S^* 
	&=&  \frac{\pi \ell^2_4}{2G_4}\left( \sqrt{ 1  + \frac{16G^2_4Q^{*2}}{\ell^2_4 }}  - 1 \right) = \pi k \left( \sqrt{ 1  + 4 k^{-2} (Q^* \ell_4)^2}  - 1 \right) ~,
	\label{eqn:BPSentropy}
\end{eqnarray}
where the omnipresent dimensionless ratio
\begin{equation}
\label{eq:kdef}
k = \frac{\ell^2_4}{2G_4} = \frac{\sqrt{2}}{3} N^{\frac3{2}}  ~,
\end{equation}
is a large pure number that sets the scale for the conserved charges. It quantifies that the black hole is much bigger than the Planck scale
with a precise value that is characteristic of the microscopic theory. The second equality applies when the AdS$_4$ background arises from $N$ $M2$-branes, 
or from their dual description by ABJM theory.

%%%%%%%%%%%%%%%%%%%%%%%%%
\subsection{NearBPS Thermodynamics}

In this subsection we initiate our
study of thermodynamics in the nearBPS regime.
The parametric representation of the BPS limit is $q=q^*$ and $r_+=r^*$ 
so we define nearBPS black holes as those where 
\begin{equation}
q-q^*\sim r_+-r^*\sim\epsilon~,
\label{eq:qrsmall}
\end{equation}
are small. The identity \eqref{eqn:mexact} shows that this is possible only if $m$ is such that $m - q(1+ag)\sim\epsilon^2$. 

The black hole temperature is 
\begin{eqnarray}
	T 	&=&\frac{1}{2\pi a(1 + ag)}\left[ \frac{r_+-r_*}{\ell_4}
	\left( 1 + 6ag + a^2 g^2\right) - \frac{q-q^{*}}{\ell_4} (1 + ag)\right] 
	+ {\cal O}\left( \epsilon^2\right)~,
	\label{eqn:Tlinear}
\end{eqnarray}
at linear order in the small parameter $\epsilon$ introduced in \eqref{eq:qrsmall}. 
The nearBPS potentials are
\begin{align}
	\Phi-\Phi^* 	&= \frac{2 (q-q^*)}{q^*}  - \frac{ 2( 1 - ag )} {q^*} \left(r_+-r_*\right) 
	+ {\cal O}\left(\epsilon^2\right)~,
	\label{eqn:Philinear}
	\\
	(\Omega^*-\Omega)\ell_4 
	&=  \frac{ 2( 1 - ag )} {q^*} \left(r_+-r_*\right)+
	{\cal O}\left(\epsilon^2\right)~,
	\label{eqn:Omegalinear}
\end{align}
at the same order. It will prove advantageous to introduce a ``nearBPS potential" $\varphi$ defined by the linear combination
\begin{equation}
	\varphi \equiv ( \Phi-\Phi^*)  + (\Omega^*-\Omega)\ell_4   = \frac{2 (q-q^*)}{q^*}~.
	\label{eqn:dvarphidef}
\end{equation}
Then the conditions \eqref{eq:qrsmall} on the parameters of nearBPS black holes are equivalent to physical potentials of order
$$
T\sim\varphi\sim\epsilon~.
$$

When black holes depart from the supersymmetric limit, their mass exceeds the BPS mass $M^*$. We showed earlier that the excitation energy 
$M-M^*$ is proportional to $m-q(1 + ag)$ and the identity \eqref{eqn:mexact} established that this quantity is positive definite, 
by casting it as a sum of two squares. We now observe that the linear combination of parameters that appear in those two squares 
coincide with the temperature $T$ and the nearBPS potential $\varphi$ at linear order. Therefore, the nearBPS mass is given by the quadratic mass formula
\begin{eqnarray}
	M - M^*	&=& \frac{C_T}{2T}\left[ T^2 + \left(\frac{\varphi}{2\pi\ell_4}\right)^2\right] ~,
	\label{eq:massquad}
\end{eqnarray}
where, after collecting various proportionality factors, we find the dimensionless coefficient
\begin{eqnarray}
	\label{eqn:CTT}
	\frac{C_T}{T\ell_4} &=&  \frac{4\pi^2 \ell^2_4}{G_4} \frac{\left(ag\right)^{\frac{3}{2}}} {\left( 1 -ag\right)  \left(1 + 6ag + a^2 g^2\right) } \cr
	& = &   \frac{8\pi^2 (Q\ell_4)^3}{k^2 + 8(Q\ell_4)^2}
	~.
%	\nonumber
\end{eqnarray}
In the second form of the expression we used \eqref{eq:agBPS} and \eqref{eq:kdef} to convert the gravitational formula into a remarkably economical form that can later be compared 
with microscopic results. The notation $C_T$ in \eqref{eq:massquad} refers to the specific heat of the nearBPS black hole. The specific heat is proportional to the temperature $T$ in the nearBPS regime so the coefficient of interest is the ratio $\frac{C_T}{T}$. 

Recall that general perturbations away from the BPS locus are characterized by {\it two} variables: the temperature $T$ adds energy with charges kept fixed while the nearBPS potential $\varphi$ parametrizes deformations along the extremal surface $T=0$ that violate the 
constraint \eqref{eq:Jconstraint} on conserved charges. The specific heat refers to the first of these, 
the addition of energy through the nearBPS potential $\varphi$ is physically quite distinct. To the extent $\varphi$ can be identified with an electric potential the corresponding coefficient in the mass excess $M-M^*$ is the black hole {\it capacitance}. Interestingly, our mass formula \eqref{eq:massquad} indicates that, for the black hole studied here, the capacitance is identical to the specific heat. \footnote{They arguably differ by a factor of temperature. We highlight the parallel between the two response functions by introducing capacitance as $M-M^* = \frac{C_\varphi}{2T}\varphi^2$. In this notation $C_\varphi=C_T$.}

An analogous equality between two physically distinct linear response coefficients was previously 
noticed for nearBPS black holes in AdS$_5$ \cite{Larsen:2019oll} and in Section \ref{sec:AdS7} we 
will establish it also in AdS$_7$. The common feature of these settings is the supersymmetry breaking pattern.
The gravitational theory has (at least) ${\cal N}=2$ supersymmetry which is mildly broken by an excess energy (conjugate to temperature) or 
R-charge (conjugate to the nearBPS potential). The reasonable expectation that the corresponding symmetry breaking scales are themselves related by supersymmetry is born out in 
the ${\cal N}=2$ version of the SYK model which realizes the analogous symmetry breaking pattern in a nongravitational setting \cite{Fu:2016vas}. The nearBPS black holes 
developed in this paper offer an appropriate setting for this physical mechanism on the bulk side of the nAdS$_2$/CFT$_1$ correspondence. It would be interesting to 
further study the supersymmetry breaking pattern in supergravity. 

The constraint on conserved charges \eqref{eq:Jconstraint} can be presented as the vanishing of the ``height" function
\begin{equation}
	\label{eqn:heightdef}
	h = 2 k^{-2}  (Q\ell_4)^4 - \frac{1}{2}J^2 - J(Q\ell_4) ~.
\end{equation}
This form of the constraint makes it manifest that nonrotating ($J=0$) and uncharged ($Q=0$) black holes are both inconsistent with supersymmetry in AdS$_4$. 
In the nearBPS region we can relax the condition $h=0$.  
Any surface with constant height function $h$ is characterized by the vanishing of the differential 
\begin{eqnarray}
	\label{eqn:dhdifferen}
	dh &=&  \big[8 k^{-2}(Q\ell_4)^3-J\big] d(Q\ell_4) -(J+Q\ell_4)dJ  \cr
	&=&   \frac{k\sqrt{ag}}{\left(1 - ag \right)^3} \left[ 2ag (3+ag) d(Q\ell_4)-(1-a^2 g^2)dJ \right] ~.
\end{eqnarray}
%
%These formulae and their close relatives would %of course look more aesthetically pleasing in %microscopic units where the dimensionless %charges $Q\ell_4$ are integral. 
The constraint $h=0$ defines a line in the two-dimensional space of conserved charges $(Q\ell_4, J)$ and we can interpret nonzero values of $h$ as a coordinate along the
normal to this line, quantifying the departure from the BPS line. However, we have already introduced the nearBPS potential $\varphi$ such that $\varphi=0$ on the constraint surface and $\varphi$ is an equally good measure of the distance from the BPS surface. Indeed, the geometry of embedded surfaces
guarantees that for small values these coordinates must be proportional. 
In the following we calculate their constant of proportionality. 

In the nearBPS regime, the parameters $m$ and $q$ are proportional up to second order in $\epsilon$, as noted after \eqref{eq:qrsmall}. Therefore, at linear order
the general physical charges $Q$, $J$ given in \eqref{eq:QJM} are both proportional to $q$, with distinct proportionality factors depending $a$. 
Rescaling of $q$ with $a$ fixed therefore changes $Q$ and $J$ by a common factor. It modifies the height function as
\begin{equation}
	dh =   \left[ \big(8 k^{-2}(Q\ell_4)^3-J\big) Q\ell_4 -(J+Q\ell_4)J \right]  \frac{dq}{q^*}  = 4k^{-2} (Q\ell_4)^4\frac{dq}{q^*} ~,
\end{equation}
where we simplified using the constraint $h=0$. However, the nearBPS potential $\varphi$ 
\eqref{eqn:dvarphidef} essentially measures the scale of $q$ via $d \varphi = 2q^{-1}_* dq$ so this calculation determines the 
constant of proportionality that we seek: 
\begin{equation}
	\label{eqn:hvarphi}
	h =   2k^{-2} (Q\ell_4)^4 \varphi~.
\end{equation}
%

%%%%%%%%%%%%%%%%%%%%%%%%%
\subsection{The First Law of Thermodynamics}
We can further illuminate the nearBPS regime by explicitly verifying the first law of thermodynamics. We write it as 
\begin{equation}
dM - Td(S-S^*) =  TdS^* + (\Omega -\Omega^*) dJ + (\Phi -\Phi^*) dQ~, 
\label{eq:dMcheck} 
\end{equation}
and consider the left and right hand sides in turn. 

The quadratic formula for the nearBPS mass \eqref{eq:massquad} gives
\begin{equation}
d M =  \frac{C_T}{T} \left( T dT + \frac{1}{4\pi\ell^2_4} \varphi d\varphi \right) ~,
\label{eq:dM1stlaw}
\end{equation}
with the coefficient $\frac{C_T}{T}$ given in (\ref{eqn:CTT}). The entropy $S-S^*$ in excess of its BPS value will turn out to involve a subtlety, as we discuss below. For now we compute the
difference between the general area law \eqref{eq:ads4ent} evaluated at the respective horizon positions $r_+,r^*$, each at the same value of $a$. This procedure gives 
\begin{eqnarray}
	S-S^* &=& \frac{\pi}{G_4} \frac{(r^2_+-r^{2*})}{1-a^2 g^2} 
	\cr
	&=& \frac{\pi\ell^2_4}{G_4} \frac{1}{\left( 1 - ag\right)\left(1 + 6ag + a^2g^2\right)} 
	\left( (ag)^{\frac{3}{2}}  ~4\pi\ell_4 T + ag ( 1 +  ag) ~\varphi \right)
	~,
	\label{eqn:SminusSstar}
\end{eqnarray}
in the nearBPS regime. The linear-in-$T$ term has the correct coefficient to cancel the analogous term in the mass formula \eqref{eq:dM1stlaw} so the left hand side of \eqref{eq:dMcheck} yields an expression proportional to $d\varphi$: 
\begin{equation}
dM - Td(S-S^*) = 
\frac{1}{G_4} 
\frac{  a } {(1- ag)(1 + 6ag+ a^2 g^2)} 
\left[ 
\sqrt{ag} \varphi - \pi T\ell_4 (1 + ag)  
\right]
 d\varphi~.
\label{eq:dM1stlawL}
\end{equation}

The right hand side of the first law \eqref{eq:dMcheck} involves
the BPS entropy $S^*$. In our ``preferred" expression \eqref{eqn:BPSentropy},
it is a function of electric charge $Q$ that gives 
\begin{equation}
	\label{eqn:dSstar}
	dS^* =  \frac{8 \pi G_4 ~Q dQ}{\sqrt{1 + \frac{16G^2_4 Q^2}{\ell^2_4}}}
	= 4\pi \frac{\sqrt{ag} }{1+ag} d(Q\ell_4)~.
\end{equation}
We also need the potentials (\ref{eqn:Philinear}-\ref{eqn:Omegalinear}) recast in terms of the temperature $T$ \eqref{eqn:Tlinear} 
and the nearBPS potential $\varphi$ \eqref{eqn:dvarphidef} as
\begin{eqnarray}
\label{eq:omomstar}
	(\Omega^*-\Omega)\ell_4 
	&=&  \frac{ \sqrt{ag} (1 - ag) } {1 + 6 a g  + a^2 g^2}~ 4\pi T\ell_4
	+
	\frac{ 1 - a^2g^2 } {1 + 6 ag + a^2 g^2}~\varphi\cr
&=&	- \frac{ 2k^{-1}\ell_4 Q}{1 + 8k^{-2}(\ell_4 Q)^2 }2\pi T\ell_4
	- \frac{\sqrt{ 1 + 4k^{-2}(\ell_4 Q)^2}}{1 + 8k^{-2}(\ell_4 Q)^2}\varphi
	~,
\end{eqnarray}
and
\begin{eqnarray}
\label{eq:phiphistar}
	\Phi-\Phi^* &=&  -  \frac{ \sqrt{ag} (1 - ag) } {1 + 6 a g  + a^2 g^2}~ 4\pi T\ell_4 
	+ 	\frac{2 ag ( 3 + a g ) } {1 + 6 ag + a^2 g^2}~\varphi~\cr
	&=&	- \frac{ 2k^{-1}\ell_4 Q}{1 + 8k^{-2}(\ell_4 Q)^2 }2\pi T\ell_4
		+ \frac{ 1 + 8 k^{-2} (\ell_4 Q)^2- \sqrt{ 1 + 4k^{-2} (\ell_4 Q)^2 }}{1 + 8 k^{-2} (\ell_4 Q)^2} \varphi ~.
\end{eqnarray}
These expressions quantify the linear changes as we move off the BPS line so the terms on the 
right hand sides are equivalent to derivatives with respect to temperature $T$ and potential $\varphi$. In this subsection the formulae in terms of the intrinsic coordinate $a$ are sufficient but we record 
these results also in microscopic units for later reference. 

%so we {\it en passant} we record the thermal derivatives 
%
%\begin{eqnarray}
%\label{eqn:tempdep}
%	\frac{1}{2\pi} \partial_T\Phi  = \frac{1}{2\pi}  \partial_T\Omega\ell_4 & = & - \frac{ 2k^{-1}\ell_4 Q}{1 + 8k^{-2}(\ell_4 Q)^2 }	~,	
%\end{eqnarray}
%
%and the dependence on the nearBPS potential $\varphi$:
%
%\begin{eqnarray}
%\label{eqn:varphidep}
%		\partial_\varphi\Phi  & = &  = \frac{ 1 + 8 k^{-2} (\ell_4 Q)^2- \sqrt{ 1 + 4k^{-2} (\ell_4 Q)^2 }}{1 + 8 k^{-2} (\ell_4 Q)^2} ~,
%	\cr
%	\partial_\varphi\Omega\ell_4 & = &   - \frac{\sqrt{ 1 + 4k^{-2}(\ell_4 Q)^2}}{1 + 8k^{-2}(\ell_4 Q)^2} ~.
%\end{eqnarray}
%

Returning to our ongoing verification of the first law \eqref{eq:dMcheck}, we combine the potentials (\ref{eq:omomstar}-\ref{eq:phiphistar}) with the differential of the BPS entropy \eqref{eqn:dSstar} and find
\begin{align}
TdS^* + (\Omega -\Omega^*) dJ + (\Phi -\Phi^*) dQ  
		= & - \frac{ \varphi + \frac{4\pi T\ell_4}{1 +ag} 
		 \sqrt{ag}  } {1 + 6 ag + a^2 g^2} \left[  ( 1-a^2 g^2) gdJ - 2a g (3 + ag) dQ \right] \nonumber\\
	= & \frac{2G_4}{\ell^{3}_4}   \frac{ ( 1- ag)^3  } {1 + 6 a g + a^2 g^2} 
	\left( \frac{1}{ \sqrt{ga}}  \varphi + \frac{4\pi T\ell_5}{1 + ag}  \right)
	dh~.
	\label{eqn:firstlawv1}
\end{align} 
In the second line we took advantage of the fact that the particular linear combination of $dJ$ and $dQ$ that appears is proportional to $dh$ given in (\ref{eqn:dhdifferen}). 
Thus the relative change in the conserved charges preserves the 
``height" function \eqref{eqn:heightdef} $h={\rm constant}$, for example by remaining within the constraint surface $h=0$.
Because of this property, we can invoke
(\ref{eqn:hvarphi}) and rewrite the differential in terms of the nearBPS potential $\varphi$
\begin{equation}
TdS^* + (\Omega -\Omega^*) dJ + (\Phi -\Phi^*) dQ 
= \frac{1}{G_4} 
\frac{  a^2g } {(1- ag)(1 + 6ag+ a^2 g^2)} 
\left( \frac{1}{ \sqrt{ag}}  \varphi + \frac{4\pi T\ell_5}{1 + ag}  \right)
 d\varphi~.
	\label{eqn:firstlawfinal}
\end{equation}
The first law of thermodynamics \eqref{eq:dMcheck} demands that this expression agrees with \eqref{eq:dM1stlawL}. The fact that both are proportional to $d\varphi$ shows that temperature changes $dT$ match, as they should. The coefficients of $\varphi d\varphi$ also coincide
but the terms proportional to $Td\varphi$ do {\it not} agree. The reason, previously uncovered in the analogous 5D setting \cite{Larsen:2019oll}
(and alluded to as a subtlety earlier in this subsection), is a dependence on the reference point on the BPS surface. 

The BPS entropy $S^*$ is defined only modulo the constraint $h=0$ on the charges. Therefore expressions that are equivalent when $h=0$ is imposed may have differentials that differ by $dh$. Indeed, the BPS entropy $S^*$ employed as reference when computing $S-S^*$ in \eqref{eqn:SminusSstar} is the general area law \eqref{eq:ads4ent}, evaluated at the BPS point. 
In contrast, the differential of $S^*$ was derived in \eqref{eqn:dSstar} from the ``preferred" form of the entropy \eqref{eqn:BPSentropy}, expressed in terms of the physical charge $Q$. 
The former amounts to a formula depending entirely on the coordinate $ag$ along the BPS line, but the latter also takes into account that $Q$ is proportional to $q$. This 
amounts to an additional contribution:
$$
\delta dS^* = 4\pi \frac{\sqrt{ag} }{1+ag} (Q\ell_4)\frac{d(q-q^*)}{q^*} = \frac{\pi\ell_4}{G_4}  \frac{ag}{1- a^2g^2}  d\varphi ~.
$$
Adding this expression to \eqref{eqn:firstlawfinal} we recover 
\eqref{eq:dM1stlawL}, as required by the first law. 

The quantitative output of this subsection is the evaluation of the
entropy due to the violation of the constraint. Its value read off from \eqref{eqn:SminusSstar} at $T=0$
\begin{eqnarray}
\label{eqn:CET}
\partial_\varphi S  &=& \frac{\pi\ell^2_4}{G_4} \frac{ag ( 1 +  ag) }{\left( 1 - ag\right)\left(1 + 6ag + a^2g^2\right)} 
= 2\pi k^2 \frac{ \sqrt{ k^2 + 4(Q\ell_4)^2}}{ k^2  + 8(Q\ell_4)^2}~,
\end{eqnarray}
defines a third response coefficient, above and beyond the two implied by the quadratic mass formula \eqref{eq:massquad}. In view of the ambiguity discussed above we must specify 
that the differentiation in its definition is taken at fixed value of the intensive parameter $a$ which, as for Kerr-Newman black holes in asymptotically flat space, equals the
ratio of physical variables $J/M$ .

%%%%%%%%%%%%%%%%%%%%%%%%%
\subsection{Stability and Physical Conditions}

The potential $\varphi$ was introduced in \eqref{eqn:dvarphidef} as a linear function of $\Phi$ and $\Omega$ that vanishes for BPS black holes and measures
the departures from the BPS line that preserve extremality $T=0$. We see from \eqref{eqn:SminusSstar} that it was defined such that the entropy {\it increases} for 
$\varphi\geq 0$. This inequality suggests that the physical configuration space is restricted to $\varphi\geq 0$. An equivalent statement is that the 
constraint relating physical charges $Q,J$ can be violated, but only such that the height function introduced in \eqref{eqn:heightdef} is positive $h\geq 0$. 
Yet another version of the inequality is that the charge parameter $q\geq q^*$.

For a perspective on these conditions we analyze Gibbs' free energy. Starting from the on-shell action \eqref{eq:onact} we can write it as
\begin{equation}
G(T, \Omega, \Phi)   =  - \frac{1}{4G_4} 
\frac{r^2_++a^2}{\Xi r^3_+}
\left[  \left(  r^4_+ g^2- a^2\right) + \frac{1}{4}( \Phi^2 - \Phi^2_* ) \left( r^2_+- a^2 \right) 
\right] ~,
\label{eqn:Gibbsfree}
\end{equation}
where the horizon position $r_+$ and the angular momentum to mass ratio $a$ are interpreted as functions of the potentials $T, \Omega$. These functions are determined implicitly through
$$
1 - \Omega^2 \ell^2_4 = \frac{ \left( r^4_+ - a^2\ell_4^2\right)\left( 1 - a^2g^2\right)}
{\left( r^2_+ + a^2\right)^2}~,
$$
and
$$
4\pi T = \frac{r_+}{r^2_++a^2} \left[ \left( r^4_+ g^2 - a^2\right) \frac{1}{r^2_+} \left( 3 + \frac{a^2}{r^2_+}\right) 
-  \frac{1}{4}( \Phi^2 - \Phi^2_* ) \left( 1 + \frac{a^2}{r^2_+}\right)^2\right] ~.
$$

We want to examine the range of parameters that corresponds to physical black holes. We first demand  that the extensive variables mass $M$, 
angular momentum $J$, charge $Q$, and entropy $S$ are finite and non-negative. This restricts the parameters so $0\leq ag<1$. The inequality 
$a\geq 0$ (and so $\Omega\geq 0$) does not limit generality because $a\to -a$ leaves all thermodynamic 
formulae (and the entire geometry) invariant, except for a flip of parity. Our second physical requirement is that $\Omega\ell_4\leq 1$, in order that the speed of light is not exceeded in the dual boundary theory. Since we already took $|ag|<1$ this condition amounts to 
\begin{equation}
\label{eqn:r2a}
r^2_+ g^2 \geq  a g \geq 0~.
\end{equation}
Interestingly, the first inequality in \eqref{eqn:r2a} can be recast as $r_+\geq r_*$ so the nonBPS black holes are all {\it larger} than their BPS relatives when measured in the conventional $r$ coordinate.

Our goal is ultimately to describe nearBPS black holes as excitations of BPS black holes. In the grand canonical ensemble considered here such states can be reached by deforming the potentials $(\Phi, \Omega)$ away from their BPS values $(\Phi_*, \Omega_*)$ while staying at extremality $T=0$, followed by raising the temperature while keeping $(\Phi, \Omega)$ fixed. This motivates our third physical condition: the potentials $(\Phi, \Omega)$ must be consistent with extremality $T=0$. 
Given the general restrictions \eqref{eqn:r2a} already imposed, vanishing temperature is possible only for $\Phi^2 \geq\Phi^2_*=4$. 
This leaves a physical domain defined by 
\begin{equation}
1\geq \Omega\ell_4\geq 0~~;~~~\Phi\geq\Phi_*~~;~~~T\geq 0~.
\label{eqn:domain}
\end{equation}
The Gibbs' free energy \eqref{eqn:Gibbsfree} is automatically negative semidefinite in this entire region. It vanishes only for BPS black holes where
$\Omega=\Omega_*=1$, $\Phi=\Phi_*=2$. Therefore the nonBPS black holes in the entire region \eqref{eqn:domain} are stable with the proviso that the BPS black holes are only marginally stable, they can be in equilibrium with a gas of BPS particles. 

The conditions we impose may be overly strict for some purposes and there can be good reasons to relax them. On the other hand, 
the region \eqref{eqn:domain} is natural for microscopic studies. For fixed $(\Phi, \Omega)$ the temperature $T$ can be increased from zero (extremality) all the way 
to the high temperature conformal regime without any phase transitions being encountered. Specifically, the deconfined phase reigns in the entire domain, the entropy is of ${\cal O}(k)$ throughout. 

As an example that is widely studied in the literature consider the Gibbs' free energy for nonrotating black holes ($\Omega=0$). In this case it is elementary to solve the equations above explicitly. This yields the formula: 
$$
G\ell_4 = - \frac{k}{27} \left(  \sqrt{(2\pi T\ell_4)^2 + \frac{3}{4} ( \Phi^2 - \Phi^2_* )} + 2\pi T\ell_4\right)^2
\left( 2\sqrt{(2\pi T\ell_4)^2 + \frac{3}{4} ( \Phi^2 - \Phi^2_* )} - 2\pi T\ell_4\right)~.
$$
This function is manifestly smooth everywhere in the interior of the domain \eqref{eqn:domain}. 
For large temperature (at fixed $\Phi$) it takes the conformal form $G\sim - \frac{4k}{27}(2\pi T\ell_4)^3$ with the numerical coefficient familiar from studies of large AdS-Schwarzchild black holes. In particular, it appears in the hydrodynamic description that applies at large temperature \cite{Bhattacharyya:2007vs,Rangamani:2009xk,Hubeny:2011hd}. The opposite regime of small temperature is more delicate. For example, the dependence $G \sim - \frac{k}{4\sqrt{27}} ( \Phi^2 - \Phi^2_* )^{\frac{3}{2}}$
on the potential along the extremal surface $T=0$ describes small nonrotating black holes.

\begin{figure}
	\label{a0}
	\begin{center}
		\includegraphics[width=\textwidth]{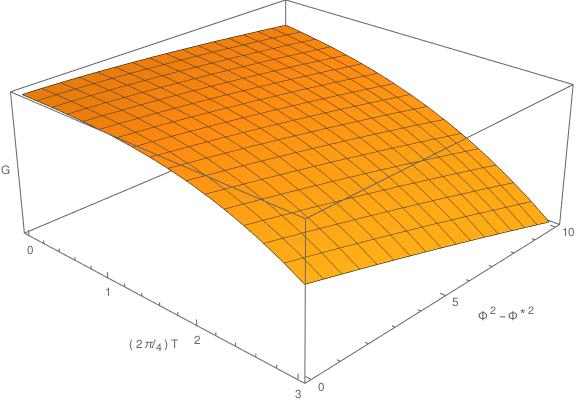}
	\end{center}
	\caption{Gibbs' free energy $G$ of the non-rotating AdS$_4$ black hole. The upper left corner is the BPS limit where $\Phi^2-\Phi^2_*=T=0$ and $G=0$. The rest of the plot has positive potentials $\Phi^2-\Phi^2_*, T$ and the free energy $G<0$. }
\end{figure}

The BPS region can be reached by tuning the potentials so $\frac{r^2_+}{a\ell_4}\to 1^+$ while simultaneously taking $\Phi\to\Phi_*^+$ with the ratio $(\Phi-\Phi_*)/(r^2_+-a\ell_4)$ kept fixed. In this limit Gibbs' free energy becomes
\begin{align}
\label{eqn:GTgen}
\frac{G}{T}  = &- (ag)^{\frac{1}{2}}k \left( \frac{ag(1+ag)}{(1-ag)^2}\frac{1-\Omega\ell_4}{T\ell_4} + \frac{\Phi-\Phi_*}{2T\ell_4} \right)~,
\end{align} 
where the dimensionless parameter $ag$ is defined implicitly by the equation 
\begin{align}
\label{eqn:aimplicit}
1 =& \frac{(ag)^{\frac{1}{2}}(3+ag)}{1-ag}\frac{1-\Omega\ell_4}{2\pi T\ell_4} - \frac{1+ag}{(ag)^{\frac{1}{2}}} \frac{\Phi-\Phi_*}{4\pi T\ell_4}~.
\end{align} 
This equation determines $a$ as a homogeneous function 
of $\Phi-\Phi^*$, $1-\Omega\ell_4$, and $T$. It is a quartic  in the variable $\sqrt{a}$ and its general solution in terms of radicals is not illuminating. 

We can ``solve" \eqref{eqn:aimplicit} for $1-\Omega\ell_4$, $\Phi-\Phi_*$ by expressing each of these quantities as a linear combination of $T$ and $\varphi$ with coefficients that depend on $a$. Such equations were already found in (\ref{eq:omomstar}-\ref{eq:phiphistar}) using other thermodynamic 
arguments and we can verify that they satisfy \eqref{eqn:aimplicit}. This is a useful consistency check. 

In the Cardy limit $1 - ag \ll 1$ we can solve the constraint \eqref{eqn:aimplicit} explicitly:
$$
1 - ag  = \frac{4 (1-\Omega\ell_4) }{g (\Phi-\Phi^*) +2 \pi  T}~,
$$
and find the free energy
$$
G = - \frac{1}{8} k \frac{(g (\Phi-\Phi^*) +2 \pi  T)^2}{g (1-\Omega\ell_4) }~.
$$
Its derivatives with respect to $T$, $\Phi$, $\Omega$ yield BPS values $S^*$, $Q^*$and $J^*$ that agree with our previous results (\ref{eqn:BPSentropy}, \ref{eq:QJM-BPS}, \ref{eq:Jconstraint}) in the Cardy limit. It is interesting that the free energy approaches the BPS limit linearly in the temperature because this reflects a BPS remnant of the familiar deconfinement transition \cite{Witten:1998zw,Aharony:2003sx,ArabiArdehali:2019orz,Cabo-Bizet:2019osg,Cabo-Bizet:2020nkr,Copetti:2020dil}.

The Gibbs' free energy \eqref{eqn:GTgen}
with $a$ given implicitly by \eqref{eqn:aimplicit}
actually describes the entire BPS surface, not just the Cardy limit. For example, we can determine the
mass as
$$
M =  \left( 1 - T\partial_T  -   \Phi\partial_\Phi  -  \Omega\partial_\Omega \right) G
= - \left(\Phi^*\partial_\Phi  + \Omega^*\partial_\Omega \right) G
= \Phi^* Q + \Omega^* J ~. 
$$
This is the exact BPS equation without assumptions on the black hole parameters. This computation is possible 
without knowing $a$ in detail because $G$ is homogeneous of degree one in the variables $(T, \Phi-\Phi^*, \Omega-\Omega^*)$ that the function $a$ depends homogeneously on. Similarly, the general derivatives of $G$ with respect to $T$, $\Phi$ and $\Omega$ depend on the unknown derivatives $\partial_T a, \partial_\Phi a, \partial_\Omega a$ but only in combinations that follow from
parametric differentiation of \eqref{eqn:aimplicit}. This procedure recovers the general expressions for  
$S^*$, $Q^*$and $J^*$ without imposing the Cardy limit. 

%%%%%%%%%%%%%%%%%%%%%%%%%%%%%%%%%%%%%%%%%%%%%%%%%%%%%%%%
\subsection{Microscopic and Macroscopic Black Hole Entropy}
\label{sec:AdS4mic}
The comparison between microscopic and macroscopic thermodynamics of BPS black holes in AdS$_4$ can be implemented conveniently by considering the 
free energy \footnote{The dimensionless numerical constant $k$ was introduced in \eqref{eq:kdef}. In this subsection we absorb the length scale $\ell_4$ into the charges $Q$ which are, therefore, quantized as integers.} \cite{Hosseini:2018dob,Zaffaroni:2019dhb}
\begin{equation}
\label{eqn:freeenergy}
\ln Z_{\rm BPS} = 4 i k \frac{\sqrt{\Delta_1 \Delta_2 \Delta_3 \Delta_4}}{\omega}~.
\end{equation}
On the microscopic side the partition function is identified with an index that can be found by methods such as supersymmetric localization. 
These computations are comparatively rigorous but the extraction of the black hole entropy from the free energy has some heuristic aspects, as we
review below. Additionally, it unclear why the index yields the black hole entropy rather than just a lower bound. 

The focus of our study is the macroscopic thermodynamics. In order to facilitate comparisons with microscopic ideas, 
we will repackage our gravitational results
into the free energy \eqref{eqn:freeenergy}. Importantly, we will do so not just for the BPS limit but for the entire nearBPS 
regime. Therefore, any microscopic computation that yields \eqref{eqn:freeenergy} accounts for the nearBPS entropy as well.

%%%%%%%%%%%%%%%%%%%%%%%%%%%%

\subsubsection{The Entropy Function}
The free energy \eqref{eqn:freeenergy} refers to the BPS partition function 
\begin{equation}
\label{eq:ZBPSdef}
Z_{\rm BPS} = {\rm Tr} \left[ e^{-\Delta_I Q^I - \omega J}\right]~.
\end{equation}
This expression applies only on the supersymmetric locus where the potentials are complex and satisfy the constraint 
\begin{equation}
\label{eqn:potconst}
(\Delta_1 + \Delta_2+\Delta_3+\Delta_4) -\omega  = 2\pi i~.
\end{equation}
We seek to compute the black hole entropy by Legendre transform of the free energy \eqref{eqn:freeenergy} to the ensemble specified by conserved charges rather than potentials. In view of the constraint \eqref{eqn:potconst}, the entropy follows from extremization of the entropy function \cite{Choi:2019zpz}
\begin{equation}
S (\Delta_I,\omega,\Lambda) = 4i k \frac{\sqrt{\Delta_1 \Delta_2 \Delta_3 \Delta_4}}{\omega} - \Delta_I Q^I - \omega J
-\Lambda (\sum_{I=1}^4 \Delta_I -\omega - 2\pi i)~,
\label{eqn:entropyfct}
\end{equation}
where $\Lambda$ is a Lagrange multiplier. The extremization conditions are
\begin{eqnarray}
\partial_{\Delta_I}S &=&  \frac{2ik}{\Delta_I} \frac{\sqrt{\Delta_1 \Delta_2 \Delta_3 \Delta_4}}{\omega} - (Q^I +\Lambda) = 0~, 
\label{eqn:extcond1}\\
\partial_{\omega}S &= & - 4i k \frac{\sqrt{\Delta_1 \Delta_2 \Delta_3 \Delta_4}}{\omega^2} - (J - \Lambda) = 0 ~,
%\cr\partial_{\Lambda}S &= &  - (\Sigma_{I=1}^4  \Delta_I -\omega - 2\pi i) = 0~.
\label{eqn:extcond2}
\end{eqnarray}
as well as the constraint \eqref{eqn:potconst}, now enforced as the equation of motion for $\Lambda$. 

Homogeneity of the free energy under rescaling of all potentials gives
$$
S - \Delta_I \partial_{\Delta_I}S - \omega\partial_{\omega}S = 2\pi i \Lambda~,
$$
so the entropy at the extremum is
\begin{equation} 
\label{eqn:S=2pi}
S = 2\pi i \Lambda~.
\end{equation} 
A combination of the extremization conditions \eqref{eqn:extcond1} and \eqref{eqn:extcond2} shows that $\Lambda$ solves the quartic equation
\begin{equation}
\label{eqn:quarticeqn}
\prod_{I=1}^4 (Q^I +\Lambda)  + k^2 (J - \Lambda)^2 = 0~.
\end{equation}
The prescription for computing the BPS entropy demands that $\Lambda$ must be purely imaginary. 
This requirement is motivated by the saddle point approximation extremizing the entropy function over complex parameters but the detailed reasoning is somewhat mysterious. As we show shortly, it has a satisfactory implication: the relation between the conserved charges it imposes is equivalent to the BPS constraint \eqref{eq:Jconstraint}. 
Moreover, after this prescription is enforced there is a unique solution with negative imaginary part, corresponding to positive entropy.

%%%%%%%%%%%%%%%%%%%%%%%%%%%%

\subsubsection{BPS Solution to the Extremization Conditions}
\label{subsec:extbps}
In view of the prescription that $\Lambda$ be purely imaginary, it is perfectly manageable to solve the quartic equation \eqref{eqn:quarticeqn} for general values of the conserved charges $(Q^I, J)$ with $I=1,2,3,4$. 
This computation yields expressions for BPS black holes that are so general that they have yet to be constructed as solutions in supergravity. 
Interesting as genericity may be, for our purposes there is some value in keeping expressions simple. We therefore focus on ``pairwise equal" charges such as  $Q^1=Q^3$, $Q^2=Q^4$. This is more general than our gravitational considerations which correspond to all charges equal. 

We take the branch $\sqrt{\Delta^2_1 \Delta^2_2} = \Delta_1 \Delta_2$ for the simplified charges and
simplify the extremization conditions \eqref{eqn:extcond1} and \eqref{eqn:extcond2} as
$$
\frac{\Delta_{1,2}}{\omega} =  \frac{Q^{2,1} + \Lambda}{2ik}~,
$$
and
$$
\frac{\Delta_1 \Delta_2}{\omega^2} = - \frac{J - \Lambda}{4ik}~.
$$
Consistency between these conditions gives the quadratic equation
\begin{equation}
    \label{eqn:quadeqn}
\Lambda^2 + (Q^1+Q^2 - i k) \Lambda + Q^1 Q^2  + i k J=0~.
\end{equation}
Its imaginary part yields
\begin{equation}
\label{eqn:sext1}
S = 2 \pi i \Lambda =  2\pi k \frac{J}{Q^1+Q^2}~.
\end{equation}
We picked the overall sign of the free energy \eqref{eqn:freeenergy} so that this entropy would be positive (for positive charges).

Recalling that $\Lambda$ is purely imaginary, the real part of the quadratic equation \eqref{eqn:quadeqn} gives
$$
\Lambda^2 - ik\Lambda + Q^1 Q^2 = 0 ~,
$$
with the solution 
\begin{equation}
\label{eqn:sext2}
S = 2\pi i \Lambda  = \pi k \left(  \sqrt{ 1 + 4k^{-2}Q^1 Q^2}  - 1 \right)~,
\end{equation}
where we chose the solution to the quadratic that gives positive entropy. 

For comparison, we recall the gravitational BPS entropy \eqref{eqn:BPSentropy}, in the case where $Q^1=Q^2\equiv Q$:
\begin{equation}
\label{eq:SeqS}
S = \pi k \frac{J}{Q}
= \pi k \left( \sqrt{ 1 + 4k^{-2}Q^2} - 1\right)~.
\end{equation}
The equality between the two forms of this gravitational formula expresses the constraint \eqref{eq:Jconstraint} satisfied by the conserved charges. The gravitational results for the BPS entropy and the constraint imposed by supersymmetry agree 
with (\ref{eqn:sext1}-\ref{eqn:sext2}) found from the extremization principle, as advertized. 

To the extent the free energy
\eqref{eqn:freeenergy} was derived from microscopic principles this provides the last step needed to arrive at the black hole entropy. Alternatively, the computation
in this subsection shows that the free energy provides a convenient packaging of the gravitational results. 

%%%%%%%%%%%%%%%%%%%%%%%%%%%%
\subsubsection{The Potentials}
The potentials $\Delta_I$ and $\omega$ introduced via the free energy \eqref{eq:ZBPSdef} and the BPS partition function \eqref{eq:ZBPSdef} are related to the gravitational potentials $\Phi$ and $\Omega$. We now proceed to compare them in detail. 

Combining the result for $\Lambda$ given in \eqref{eqn:sext2} with the constraint on the potentials \eqref{eqn:potconst}
we find (for pairwise equal charges): 
\begin{equation}
\label{eqn:compot}
\frac{2 \pi i }{\omega}  + 1 = 2\frac{\Delta_1+\Delta_2}{\omega} =  2\frac{Q^1+Q^2 + 2\Lambda}{2ik} =  - ik^{-1}(Q^1+Q^2)   -  \left(  \sqrt{ 1 + 4k^{-2}Q^1Q^2}  - 1 \right)~.    
\end{equation} 
This gives the real and imaginary parts of the potential conjugate to the angular momentum
\begin{eqnarray}
\label{eqn:Reom}
\frac{{\rm Re} ~\omega} {2 \pi }  &=& - \frac{k^{-1}(Q^1+Q^2)}{ 1 + k^{-2}[(Q^1+Q^2)^2+4Q^1 Q^2]}~,\\
\label{eqn:Imom}
\frac{{\rm Im} ~\omega} {2 \pi }  &=& - \frac{\sqrt{ 1 + 4k^{-2}Q^1Q^2}}{ 1 + k^{-2}[(Q^1+Q^2)^2+4Q^1 Q^2]}~.
\end{eqnarray}
We similarly find the real and imaginary parts of the potentials conjugate to the charges: 
\begin{eqnarray}
\label{eqn:ReDel}
\frac{4{\rm Re} ~\Delta_{1,2}} {2 \pi }  &=& - \frac{k^{-1}(Q^1+Q^2) - k^{-1}(Q^{1,2} - Q^{2,1})\sqrt{ 1 + 4k^{-2}Q^1Q^2}}{ 1 + k^{-2}[(Q^1+Q^2)^2+4Q^1 Q^2]}~,\\
\label{eqn:ImDel}
\frac{4{\rm Im} ~\Delta_{1,2}} {2 \pi }  &=&  1 - \frac{k^{-2}[(Q^{1,2})^2 - (Q^{2,1})^2] + \sqrt{ 1 + 4k^{-2}Q^1Q^2}}{ 1 + k^{-2} [(Q^1+Q^2)^2+4Q^1 Q^2]}~.
\end{eqnarray}

The potentials $\Delta_I$, $\Omega$ were introduced as the independent variables of the 
BPS partition function \eqref{eq:ZBPSdef} and determined here from extremization of the entropy function \eqref{eqn:entropyfct}.
They can not be identified with their supergravity analogues which take the values $\Phi_*=2$ and $\Omega_*=1$ identically
due to the BPS mass relation $M = 2Q + J$. To make progress we consider the general (non-BPS) partition function 
$$
Z = {\rm Tr} \left[ e^{-\beta (M - \Phi_I Q^I - \Omega J)}\right] = {\rm Tr} \left[ e^{-\beta  (\Phi_I -\Phi_{I*}) Q^I - \beta (\Omega-\Omega_*)  J)}\right]~.
$$
Comparison with \eqref{eq:ZBPSdef} suggests the identifications
\begin{eqnarray}
\label{eqn:omdeltenta1}
\omega & \overset{?}{=} & \beta(\Omega - \Omega_*)   \underset{T \to 0}{=} \partial_T \Omega~,\\
\label{eqn:omdeltenta2}
\Delta_I & \overset{?}{=} &  \beta(\Phi_I - \Phi_{I*}) \underset{T \to 0}{=}  \partial_T \Phi_I ~,
\end{eqnarray}
in the extremal limit $T\to 0$. Comparison between the expressions for $\omega, \Delta_I$ above and their analogues in gravity, \eqref{eq:omomstar} and \eqref{eq:phiphistar} establish that in fact
\begin{eqnarray}
\label{eqn:reom}
{\rm Re} ~\omega = \partial_T\Omega=&-\frac{4\pi k^{-1}Q}{1+8k^{-2} Q^2}=&
-\frac{4\pi \sqrt{a g} (1 - a g)}{1 +6a g + (a g)^2 }~,\\
\label{eqn:redelt}
4{\rm Re} ~\Delta =\partial_T\Phi=& -\frac{4\pi k^{-1} Q}{1+8 k^{-2}Q^2}=&- \frac{4\pi \sqrt{a g} (1 - a g)}{1 + 6a g + (a g)^2}~ .
\end{eqnarray}
at least when all charges $Q_I$ are identical. This establishes a natural map between the macroscopic $(\Phi,\Omega)$ and microscopic $(\Delta, \omega)$  potentials.

However, this cannot be the entire story. Physical potentials in the gravitational solution are real while the fugacities introduced in the microscopic partition function can preserve supersymmetry only if they acquire an imaginary part. The missing ingredient is the one we stress throughout this paper: the BPS surface is co-dimension two, it can be approached from two distinct directions. 

As discussed earlier, the real part of the microscopic potentials is related to increases in temperature $T$. We expect that their imaginary parts correspond to violation of the constraint, expressed in terms of potentials as $\varphi=0$.   
Indeed, comparison between the expressions for $\omega, \Delta_I$ above and their analogues in gravity
(\ref{eq:omomstar}-\ref{eq:phiphistar}) show that:
\begin{eqnarray}
\label{eqn:imome}
{\rm Im} ~\frac{\omega}{2\pi} = \partial_\varphi \Omega=&-\frac{\sqrt{\frac{4 Q^2}{k^2}+1}}{1+8k^{-2} Q^2}=&- \frac{1 - a^2 g^2}{1 + 6a g + (a g)^2}~,\\
 \nonumber
\label{eqn:imdelt}
4{\rm Im} ~\frac{\Delta}{2\pi}= \partial_\varphi \Phi= &1-\frac{\sqrt{\frac{4 Q^2}{k^2}+1}}{1+8 k^{-2}Q^2}=&\frac{2 a g (3 + a g)}{1 + 6a g + (a g)^2}~.
\end{eqnarray}
Note that while these expressions for $\omega$ and $\Delta$, expressed as functions of $a$ and $g$ exactly match the gravitational results (\ref{eq:omomstar}-\ref{eq:phiphistar}), they are ambiguous as functions of $Q$ and $J$, defined only modulo the constraint \eqref{eq:SeqS}. This is equivalent to demanding that the height above the BPS surface $h=0$. 

%%%%%%%%%%%%%%%%%%%%%%%%%%%%

\subsubsection{The nearBPS Regime}
\label{sec:ads4nearbps}
The microscopic discussion of nearBPS black holes is necessarily less rigorous than for their BPS relatives but some progress can be made nonetheless. 

A good starting point is the relation between potentials $(\Delta_I, \omega)$ in the microscopic description and their gravitational analogues $(\Phi_I,\Omega)$. Comparing the
partition functions (or the first law) at vanishing temperature gives the provisional identification (\ref{eqn:omdeltenta1}-\ref{eqn:omdeltenta2}) but supersymmetry additionally imposes the boundary conditions \eqref{eqn:potconst} on the microscopic potentials. A natural generalization is 
\begin{equation}
    (\Phi_1 - \Phi^*_1) + (\Phi_2 - \Phi^*_2) +
     (\Phi_3 - \Phi^*_3) + (\Phi_4 - \Phi_4^*) 
    -(\Omega-\Omega^*)  = \varphi + 2\pi i T \,.
    \label{eqn:potcon2}
\end{equation}
For $\varphi=0$ this boundary condition is equivalent to the BPS requirement \eqref{eqn:potconst} in the extremal limit $T\to 0$. However, for $T$ and/or $\varphi$ nonvanishing it breaks 
supersymmetry. Motivated by the success in the BPS limit, we identify the real part of the potentials $(\Phi_I,\Omega)$ in \eqref{eqn:potcon2} with their gravitational counterparts. As a practical matter, once the physical parameter $\varphi\neq 0$, the full symmetry breaking pattern is easily implemented by the substitution $\varphi \to \varphi + 2\pi i T$.

The BPS free energy \eqref{eqn:potconst} is common to 
all recent discussions of microscopic entropy for AdS$_4$ black holes \cite{Choi:2018fdc,Nian:2019pxj,BenettiGenolini:2019jdz} . A minimal framework for nearBPS statistical physics applies the modified boundary condition
\eqref{eqn:potcon2} to the BPS free energy. 
This proposal can be presented efficiently as an extremization principle for the nearBPS entropy at linear order away from the BPS surface:  
\begin{align}
TS(\Phi_I,\Omega,\Lambda)  = 
&  4i k 
\frac{\sqrt{\prod_I (\Phi_I-\Phi_I^*)} }{\Omega-\Omega^*} - \sum_I (\Phi_I-\Phi^*_I) Q^I - (\Omega-\Omega^*) J\nonumber\\
& - \Lambda (\sum_{I=1}^4 (\Phi_I-\Phi_I^*) -(\Omega-\Omega^*) - \varphi - 2\pi i T )~.
\label{eqn:nBPSentFct}
\end{align}
We do not derive our prescription {\it ab initio}, but it is arguably a corollary of previously accepted microscopic considerations.
It is thought that the BPS index can be continued freely from weak to strong coupling and, additionally, that the index and the partition function have the same asymptotic behavior in the gravitational regime. Any agreement in the strict BPS limit relies on these features and the only additional ingredient we invoke is smoothness of gravitational thermodynamics as the nearBPS regime approaches the BPS limit. An even more conservative view is that agreements we establish in the following show that our nearBPS extremization principle provides an efficient packaging of gravitational data.  

It is straightforward to make our proposal explicit for four generic charges but, as in subsection \ref{subsec:extbps}, we prioritize transparency over generality and focus on the case where charges are equal in pairs and expressions are more illuminating. Then the values for the potentials at the extremum differ from the BPS results (\ref{eqn:Reom}-\ref{eqn:ImDel}) only by some simple substitutions. We write them as
%
%\begin{eqnarray}
% ~\frac{\Omega- \Omega^*}{\varphi + 2 \pi i T}   
% &=& \frac{-\sqrt{1 + 4k^{-2}Q^1 Q^2} + i k^{-1} (Q^1+Q^2)}{1 + k^{-2} ((Q^1+Q^2)^2+ 4Q^1 Q^2) }
%\cr
%\end{eqnarray}
%
%
%\begin{eqnarray}
%2 \frac{\Phi_{1,2}- \Phi_{1,2}^*}{\varphi + 2 \pi i T} 
% &=& 1 - 
% \frac{   \sqrt{1 + 4 k^{-2} Q^1 Q^2} + k^{-2}((Q^1)^2-(Q^2)^2) + i k^{-1}[ (Q^{1,2}-Q^{2,1}) \sqrt{1 + 4 k^{-2} Q^1 Q^2}
%  -(Q^1  + Q^2)]}{
% 1 +  k^{-2} ((Q^1+Q^2)^2 + 4 Q^1 Q^2)}\nonumber
%~.
%\end{eqnarray}
\begin{eqnarray}
 ~\frac{\Omega- \Omega^*}{\varphi + 2 \pi i T}   
 &=& - \frac{1}{\sqrt{1 + 4k^{-2}Q^1 Q^2} 
 + ik^{-1} (Q^1 + Q^2)  }~,
\cr
%\end{eqnarray}
%
%
%\begin{eqnarray}
4 \frac{\Phi_{1,2}- \Phi_{1,2}^*}{\varphi + 2 \pi i T} 
 &=& 1 - 
 \frac{  1  + i k^{-1} (Q^{1,2}-Q^{2,1})}{
 \sqrt{1 + 4 k^{-2} Q^1 Q^2} + ik^{-1} (Q^1 + Q^2))}\nonumber
~.
\end{eqnarray}
After multiplication with $\varphi + 2 \pi i T$ on both sides of the equations, the real part of each potential becomes a linear combination of $\varphi$ and $T$. These expressions agree with the analogous results computed from the black hole solutions (\ref{eq:omomstar}-\ref{eq:phiphistar}). This result streamlines the identifications we already reported in (\ref{eqn:reom}-\ref{eqn:imdelt}) by incorporating them in a systematic computation. 

The nearBPS extremization conditions are identical to their BPS counterparts  (\ref{eqn:extcond1}-\ref{eqn:extcond2}), except for simple substitutions of variables. We do not need the details because the quartic equation in the Lagrange multiplier $\Lambda$ \eqref{eqn:quarticeqn} with coefficients depending on charges $(Q^I, J)$ is not modified. 
The important new feature is that solutions to the quartic where $\Lambda$ is purely imaginary are insufficient.
We insert the more general root in the on-shell 
nearBPS entropy function  
\begin{align}
\label{eqn:Freedef}
    TS &= \Lambda ( \varphi + 2\pi i T )\,,
\end{align}
and identify the resulting real part as
the physical entropy. This gives a corrected value for the entropy.  

Since we consider charges that are equal in pairs the quartic equation satisfied by $\Lambda$ \eqref{eqn:quarticeqn} reduces to the quadratic \eqref{eqn:quadeqn} which we recast as
\begin{align}
    \left(\Lambda +\frac{i J k}{Q_1 + Q_2}\right) \left(\Lambda-i k\big(\frac{J}{Q_1 + Q_2}+1\big) +Q_1 + Q_2\right)= -\frac{2 k^2 h}{(Q_1+Q_2)^2}\,,
\end{align}
where the ``height" function 
\begin{align}
h= \frac{1}{8}\left(  1 + 4 k^{-2} Q_1 Q_2\right) (Q_1 + Q_2)^2
- \frac{1}{2} \left( J  +  \frac{Q_1+Q_2}{2}\right)^2\,,
\end{align}
generalizes $h$ introduced in \eqref{eqn:heightdef} to permit two distinct charges. This form of the equation for $\Lambda$ makes it manifest that, when the constraint on charges
$h=0$ is imposed, we have purely imaginary $\Lambda$ with the value given in \eqref{eqn:sext1} modulo any rewriting using $h=0$. 

Conversely, when we allow violation of the constraint between charges by taking non-zero $h$ the Lagrange multiplier is shifted. At linear order we find
\begin{equation}
\label{eqn:delLambda}
\delta\Lambda 
= \frac{2 k h}{(Q^1+Q^2)^2\left[i\sqrt{ 1 + 4k^{-2} Q_1 Q_2}  -  k^{-1}(Q^1 + Q^2)\right]}~.
\end{equation}
Since this result is already proportional to $h$ we freely applied the constraint $h=0$ to eliminate $J$ from the coefficient. 

For comparison with gravity, we relate the height-function $h$ to the potential $\varphi$ through the generalization of  \eqref{eqn:hvarphi} to two independent charges
$$
h = \frac{1}{2}k^{-2} (Q_1+Q_2)^2 Q_1 Q_2 \varphi~.
$$
The prescription $\varphi \to \varphi + 2\pi i T$ then gives 
\begin{align}
S - S^* =& {\rm Re} ~[2\pi i \delta\Lambda] 
%= {\rm Re}~ [ 2\pi i \frac{\varphi + 2\pi i T}{\left[i\sqrt{ 1 + 4k^{-2} Q^1 Q^2} \right) -  k^{-1}(Q^1 + Q^2)]}Q^1Q^2]
\nonumber\\
=& 2\pi  k Q^1Q^2 \frac{\sqrt{ 1 + 4k^{-2} Q^1 Q^2} ~\varphi 
+ 2\pi (Q^1 + Q^2)~T }{1 + k^{-2} [ 4Q^1 Q^2  +  (Q^1 + Q^2)^2]}~.
\end{align} 
This agrees precisely with the gravitational result \eqref{eqn:SminusSstar} after 
specialization to equal charges $Q=Q_1=Q_2$ and then trading $Q$ for 
the gravitational parameter $a$ via \eqref{eq:QJM-BPS}. The two variables being compared are both linear response coefficients, related to the the specific heat \eqref{eqn:CTT} and the electric field \eqref{eqn:CET}, respectively. Thus our result relates parameters of nonsupersymmetric black holes to microscopic concepts.  

The agreement reported in this subsection goes against expectations from rigid versions of indexology that demand strict adherence to supersymmetry. However, it is less surprising from an effective quantum field theory point of view. 
It is expected that the UV theory accounts for the supersymmetric ground state entropy, i.e. the size of the classical phase space at very low energy. The leading excitations above the ground state are described by a low energy effective field theory with gravitational/QFT aspects encoded in the  nAdS$_2$/CFT$_1$ 
correspondence \cite{Almheiri:2016fws,Sachdev:2019bjn}. 
It depends on just a few symmetry breaking parameters that generally must be determined by matching to the UV theory. Although we have not developed the effective theory systematically, it is not unreasonable that we can recover these effective parameters quantitatively by studying collective modes on the agreed upon classical phase space. 

%%%%%%%%%%%%%%%%%%%%%%%%%%%%%%%%
\section{Asymptotically AdS\texorpdfstring{$_7$}{7} Black Holes}
\label{sec:AdS7}

In this section we study the near BPS thermodynamics of black holes in AdS$_7$. The maximally supersymmetric theory results from eleven-dimensional supergravity reduced on S$^4$. As a result, the quantum numbers of a generic black hole solution in this theory are the mass $M$, three angular momenta $(J_1, J_2, J_3)$ that correspond to rotations in AdS$_7$, and two charges $(Q_1, Q_2)$ that correspond to momenta along $S^4$. A completely general solution has not yet been constructed but a special case with three identical angular momenta and two independent charges was first presented in \cite{Cvetic:2005zi}. In the gravitational part of our calculations we consider a further simplification to the Kerr-Newman AdS$_7$ black hole, i.e. the special case where the two charges are equal. This geometry is also a solution to minimal supergravity in AdS$_7$. 

%%%%%%%%%
\subsection{The Black Hole Geometry}
The conserved charges $(M, J, Q)$ are encoded in a mass parameter $m$, an angular momentum parameter $a$, 
and a charge parameter $q$ (which we occasionally trade for the ``boost" parameter $\delta$ introduced through $q=m\sinh^2\delta$). 
The geometry presented in \cite{Cvetic:2005zi} is 
%%%%
\begin{align}
ds_7^2 &= (H)^{2/5}\, \Big[ -\frac{Y\, dt^2}{f_1\, \Xi_+^2}
+ \frac{\rho^6\, d\rho^2}{Y}+ \frac{f_1}{\rho^4\, H^2 \, \Xi^2}\,
\Big(\sigma + \frac{2 f_2}{f_1}\, dt\Big)^2 +
\frac{\rho^2}{\Xi}\, d\Sigma_2^2\Big]\,,
\end{align}
where
\begin{align}
H &= 1 + \frac{2\,q}{\rho^4}\,,&& \rho^2= r^2+a^2\,,\nonumber\\
\Xi_\pm &\equiv 1 \pm a\, g\,,&& \Xi \equiv  1 -a^2\, g^2
= \Xi_-\, \Xi_+\,,\nonumber
% \\q&= m\sinh^2\delta\,, && c=\cosh\delta\,,\, s=\sinh\delta\,,
\end{align}
%%%%%
and the functions $f_1$, $f_2$ and $Y$ are given 
%%%%%
\begin{align}
f_1 &= \frac{\Xi}{\rho^2}(\rho^4+2q)^2 +
2 a^3\, g\, q\, (2-a\, g)-
\frac{4\Xi_-^2\, a^2\, q^2}{\rho^4}+2\, m\, a^2\,,\nonumber\\
f_2 &=  \frac{g\Xi_-}{2\rho^2} \Big(\rho^4+2q\Big)^2+a(q+m)\,,\nonumber\\
Y 
%&= (r^2+a^2)^3(1+r^2g^2)+4\,q\,g^2(r^4+q) + 2 a^2\,g\, q\, (3\,g\,r^2+2\,a)-2\, m\,r^2\,\nonumber\\
&=g^2 \rho^8+\Xi \rho^6+4 g^2 \rho^4 q-2\left(m+a^2 g^2 q\right)\rho^2+4 g^2 q^2+2 a^3 g q (2-ag)+2 m a^2\,.
\label{eq:Ydef}
\end{align}
%%%%%
The $d\Sigma^2_2$ is a metric on $\mathbb{CP}_2$ which together with the $U(1)$ fibre $\sigma$ forms a five-sphere within AdS$_7$. The omnipresent constant $g$ is the coupling constant of gauged supergravity, related to the radius of the asymptotically AdS$_7$ spacetime as $g=\ell_7^{-1}$. 
For $ag$ outside the 
range $(-1,1)$, the coefficient of $d\Sigma_2^2$ is nonpositive (or divergent) so the spacetime signature is not Lorentzian. 

The event horizon of the black holes is located at the coordinate $r=r_+$ where the function $Y(r)$ has its largest root. The thermodynamic 
potentials characterizing the solution are all evaluated at this value $r=r_+$. They 
are:\footnote{We believe there are misprints in the thermodynamic quantities given in \cite{Cvetic:2005zi} and \cite{Choi:2018hmj,Hosseini:2018dob}. 
	Our expressions satisfy a number of consistency checks and agree with those given in \cite{Bhattacharyya:2007vs}.}
\begin{align}
\label{eq:ST-7}
\frac{4G_7}{\pi^2} S &=\frac{\pi\, \rho^2_+ \, \sqrt{f_1(r_+)}}{\Xi^3}\,,\nonumber\\
T &=\frac{\left.\partial_r Y\right|_{r=r_+}}{4 \pi\,  r_+(r^2_++a^2)\, \sqrt{f_1(r_+)}}=\frac{4 g^2 \rho^6_++\,3 \Xi \rho^4_++8 g^2\rho^2_+ q-2 a^2 g^2 q-2 m}{2 \pi  \rho^2_+\sqrt{ f_1(r_+)}}\,,\nonumber\\
\Omega &= \frac{2f_2(r_+)\, \Xi_+}{g f_1(r_+)} -1= 
2a 
\frac{\rho^4_+ \left(m +q (1+ag-2a^2g^2+a^3g^3)\right) + 2a g q^2\Xi_-^2}{g\, \rho^4_+ f_1(r_+) }
\,,\nonumber\\
\Phi&=\frac{2m\sinh 2\delta}{(\rho^4_++2q)\, \Xi_-}\, \left[1-\frac{2a f_2(r_+)\, }{f_1(r_+)}\right]\,.
\end{align}
%%%%%

The conserved charges of the black holes are given in terms of the parameters $(m, a, q=m\sinh^2\delta)$ through
%%%%%
\begin{align}
\frac{4G_7}{\pi^2} ~M &=\frac{1}{2\Xi^4}  \left[m\left(6-\Xi\right) - q\left(3\Xi_-^4-7\Xi_-^2+2\Xi_--6\right)\right]\,, \nonumber\\
\frac{4G_7}{\pi^2} ~J &= \frac{1}{\Xi^4}  a \left[m-q \left(\Xi_-^3-\Xi_-^2-1\right)\right]\,,\nonumber
%=\frac{1}{\Xi^4}  \left[\left(1-\Xi_-\right)m+q \left(\Xi_-^4-2\Xi_-^3+\Xi_p^2-\Xi_p+1\right)\right]\,\nonumber,
\\
\label{eqn:MJQ}
\frac{4G_7}{\pi^2} ~Q &= \frac{1}{2\Xi^3} m\, \sinh2\delta\,.
\end{align}
The length dimensions of the variables in these formulae are $[G_7]=L^5$, $[m]=[q]=L^4$, and $[a]=L$. The
physical variables therefore have dimensions $[J]=L^0$, $[M]=[Q]=L^{-1}$, in agreement with expectations from supergravity. 

In the context of the AdS/CFT-correspondence, the gravitational coupling constant $G_7$ in units of the AdS$_7$ radius $\ell_7=g^{-1}$ is 
related to the integral $M5$-brane number $N$ via 
\begin{equation}
\label{eq:G7N3}
\frac{\ell^5_7}{G_{7}}  = \frac{16}{3\pi^2}  N^3~.
\end{equation}
In the holographically dual CFT the dimensionless mass $M\ell_7$ is the conformal dimension, $Q\ell_7$ is the quantized $R$-charge, and $J$ is the 
angular momentum quantized in units of $\hbar=1$. These quantum numbers are all macroscopic ${\cal O}(N^3)$ for a black hole with parameters $m, q, a^4$ of order ${\cal O}(\ell^4_7)$.  

%%%%%%%%%%%%%%%%% 
\subsection{Supersymmetric Black Holes}
Extremal black holes have zero temperature. Referring to the expression for temperature (\ref{eq:ST-7}), this is equivalent to the derivative $\partial_r Y=0$ at the event horizon $r=r_+$. Since the polynomial $Y(r)$ also vanishes there, it develops a double root. The extremality condition $\partial_r Y(r_+)=0$ gives a simple equation for the mass parameter $m$
\begin{align}
\label{eq:mextremal}
m= 2 g^2 \rho_+^6+\frac32 \Xi \rho_+^4+4 g^2 q\rho_+^2- a^2 g^2 q\,.
\end{align}
Combining this with the horizon condition $Y(r_+)=0$
where $Y$ is given in \eqref{eq:Ydef} yields an equation for the location  $r_+$ of the horizon in terms of black hole parameters $a$ and $q$.
This in turn yields the mass parameter $m$ through (\ref{eq:mextremal}). Thus the extremality condition $T=0$ defines a relation between the physical black hole variables $(M, J, Q)$. It can be interpreted as giving the lowest possible mass $M=M_{\rm ext}$ for given conserved charges $J,Q$. 

While this procedure is straightforward in principle, in practice it is unwieldy and not terribly illuminating. To make progress we therefore take supersymmetry into account. For the theory to admit supersymmetric solutions, all physical configurations must satisfy the BPS bound
\begin{align}
\label{eq:BPSmass}
M-3 g J - 4 Q\ge 0\,, 
\end{align}
with equality for BPS black holes. The linear combination of conserved charges that appear can be recast as
\begin{align}
\label{eqn:MnMstar}
M-3 g J - 4 Q =& -\frac{m\,\pi^2}{32G_7}\frac{3\, \Xi_- a g (2-a\,g)\left(1+3 a\,g\right)}{e^{2\delta}\Xi^4} \left( e^{2\delta}-1 - \frac2{3\,a\,g}\right)\times\nonumber\\
&\hspace{5.9cm}\left( e^{2\delta}-1+2\frac{5-a\,g}{(2-a\,g)\left(1+3 a\,g\right)}\right).
\end{align}
The expression in the parenthesis on the second line is strictly positive in the entire physical range $0\leq ag<1$ so the BPS bound amounts to:  
\begin{align}
\label{eq:delta-BPS}
1+\frac2{3\,a\,g} \, \geq\, e^{2\delta}   \,.
\end{align}
An alternative expression follows from the identity
	\begin{align}
	(M-3gJ)^2-(4Q)^2=&\frac{\pi^4\Xi_-^2}{64\,G_7^2\, \Xi^8}\left(m -  3 a g(2+ 3 a g)  q \right) \times\nonumber\\
	&\left(m(5- a g)^2 + q(2-a g) (1+ 3 a g)(8-7 ag + 3 a^2 g^2)\right)~,
	\end{align}
	which yields the parametric form of the BPS bound
	\begin{align}
	\label{eq:qmBPS}
	m \ge  3 a g(2 + 3 a g) q \,.
	\end{align}
The equivalence of this inequality and \eqref{eq:delta-BPS} is easily verified using the definition $q=m\sinh^2\delta$. Both inequalities are saturated if and only if the black hole is supersymmetric. 

BPS saturation is possible only for extremal black holes but it is a stronger condition, it imposes a constraint on the black hole parameters in addition to vanishing temperature.
Some authors impose this ``second" condition by demanding the absence of closed time-like curves, a requirement that 
makes reference to detailed analysis of the geometry. Later in this subsection 
we show that BPS saturation automatically gives both vanishing temperature and the additional constraint on black hole parameters. Thus the latter does not require appeal to an independent physical principle. 

In preparation for this argument we temporarily impose the BPS formula for the mass and independently set the temperature to zero. Accordingly, we assume that $m$ and $q$ are related by 
equality in \eqref{eq:qmBPS} and additionally require that $m(r_+)$ is the function given in \eqref{eq:mextremal}. In this situation the horizon equation $Y(r_+)=0$ becomes
\begin{align}
Y(r_+)&=\frac{4\, g^2 \rho_+^4 (1+3 a g)^2 \left(\rho_+^2-\frac{4\,a^2\Xi_+}{1+3 a g}\right)^2 \left(\rho_+^2+\frac{3 \left(3+5 a g\right)\Xi_-}{16 g^2}\right)}{\left(2 g\rho_+^2-3 a-5 a^2 g\right)^2}=0\,.
\end{align}
It is manifest that the largest root is a double root, as expected, and locates
the event horizon at 
\begin{align}
\label{eq:BPSradius}
\rho^*_+= 2 a \sqrt{\frac{ 1 + a g }{1+3 a g}}\,.
\end{align}
Here and in the following we use the superscript $^*$ to denote quantities that take on their BPS values. The location of the event horizon 
\eqref{eq:BPSradius} gives the BPS values 
\begin{align}
m^*&=\frac{12 a^4 (1+a g)^3 (2+3 a g)}{(1+3 a g)^2}\,,\\
\label{eqn:qstar}
q^*&=\frac{4a^3 (1+ag)^3}{g (1+3 a g)^2}\,.
\end{align}
Our BPS values for these parameters agree with those
found in \cite{Cvetic:2005zi}.
\footnote{However, our value for $\rho_+$ in \eqref{eq:BPSradius} differs from the one given in \cite{Cassani:2019mms}.}
They correspond to the physical quantum numbers: 
\begin{align}
\ell_7 M^*&= N^3 \frac{16a^3 g^3 \left(4+11 a g+6 a^2 g^2+3 a^3 g^3\right)}{ 3(1-a g)^4(1+3 a g)^2}\,,\nonumber\\
J^*&= N^3 \frac{16a^4 g^4 \left(1+6 a g+a^2 g^2\right)}{3(1-a g)^4(1+3 a g)^2}\,,\nonumber\\
\ell_7 Q^*&= N^3 \frac{16a^3 g^3}{3(1-a g)^3(1+3 a g)}\,,
\label{eq:EJQbps}
\end{align}
and the BPS black hole entropy becomes
\begin{align}
\label{eq:sbps}
S^*&=2\pi N^3 \frac{16a^4 g^4}{3(3+ag) (1-ag)^3} \left(\frac{3+ag}{1+3ag}\right)^{3/2}\\
&= 2 \pi  \sqrt{\frac{6 (\ell_7 Q^*)^3- 3 N^3 (J^*)^2}{6 \ell_7 Q^*-N^3}}~.
\label{eq:BHentropy}
\end{align}
In these BPS formulae we have opted to express the overall normalization in microscopic units via \eqref{eq:G7N3}. 

The physical variables satisfy the BPS mass formula $M^*=3gJ^*+4Q^*$, as they should. 
In our conventions the first law of thermodynamics takes the form
\begin{align}
\label{BPS1stlaw}
dM -2\Phi dQ -3\Omega dJ =  T dS   \,,
\end{align}
so on the extremal surface $T^*=0$, the BPS mass formula gives
\begin{align}
\Omega^*&=g\,,\nonumber\\
\Phi^* &=2\,.\nonumber
\end{align}
The general thermodynamic potentials \eqref{eq:ST-7} do in fact simplify to 
these constants when they are evaluated on the BPS surface. 
It is an additional consistency check on our formulae that the first law (\ref{BPS1stlaw}) is satisfied after expressing each of the BPS quantities as functions of $a$.

The BPS expressions given above were all computed assuming both saturation of the BPS bound \eqref{eq:qmBPS} and vanishing temperature, as implemented through \eqref{eq:mextremal}. However, with the benefit of hindsight we can now do better. We begin with rewriting the exact metric function $Y$ as a power series in $\rho^2$ around the position of the BPS horizon: 
\begin{align}
    Y(r)=&-2 r^2\left(m-3ag(2+3ag)q\right)+4g^2 \left(q-q^*\right)^2 -\frac{4 a g  (3 + 6 ag + 7 a^2g^2)}{1 + 3 a g} \left(\rho^2-\rho^{*2}_+\right) \left(q-q^*\right) \nonumber\\
    &+\frac{4 a^2  (1 + a g)^2 (3 + 10 ag + 19 a^2 g^2)}{(1 + 3 a g)^2}\left(\rho^2-\rho^{*2}_+\right)^2 +4g^2 \left(\rho^2-\rho^{*2}_+\right)^2\left(q-q^*\right)  \nonumber\\
    &+\frac{(1 + a g) (1 + 2 ag + 13 a^2 g^2)}{1+3 a g}\left(\rho^2-\rho^{*2}_+\right)^3 +g^2\left(\rho^2-\rho^{*2}_+\right)^4
   ~.
\end{align}
The horizon equation $Y(r_+)=0$ then gives
\begin{align}
\label{eqn:quadmq}
   &  2 r_+^2\left(m-3ag(2+3ag)q\right) = 
   \frac{g^2(3 + ag)(1 + 3ag)^3(q-q^*)^2}{(1+a g)^2 (3 + 10 ag + 19 a^2g^2)} 
   \\
& +   \sqrt{ 3 + 10 ag + 19 a^2g^2}
    \left[ \frac{2 a   (1 + a g) }{1 + 3 a g}\left(\rho^2_+-\rho^{*2}_+\right) - \frac{3 + 6 ag + 7a^2g^2}{(3 + 10 ag + 19a^2g^2)(1 + ag)}(q-q^*)g \right]^2 ~, \nonumber
\end{align}
near the BPS limit.\footnote{Our computations here are valid up to quadratic departures from the BPS limit. The analogous formulae for AdS$_4$ and AdS$_5$ black holes can be made exact without much additional effort. We assume that the result is exact in AdS$_7$ as well but we have not worked out the details, as they are more elaborate in this case. 
} The left hand side of this expression is non-negative and vanishes exactly when the BPS bound (\ref{eq:qmBPS}) on the mass is saturated.
Since the right hand side is manifestly the sum of two non-negative terms, we see that BPS saturation implies {\it two} conditions on the black hole. Moreover, the large square bracket on the second line is proportional to the temperature
\begin{align}
\label{eqn:linearT}
T = \frac{g}{2\pi a^2} \frac{1}{1+ag}\sqrt{\frac{1+3ag}{3+ag}}
\left[ 
\frac{3 + 10 ag + 19a^2g^2}{1+3ag}\left(\rho^2_+-\rho^{*2}_+\right) - 
\frac{3 + 6 ag + 7a^2g^2}{2 a(1+ag)^2}(q-q^*)g
\right]~,
\end{align}
at linear order. Therefore, one of the two conditions is extremality $T^*=0$. The other requirement is the constraint on potentials for conserved charges, identified here in the form $q=q^*$. 

In view of the two independent conditions on the black hole parameters that follow from supersymmetry, BPS black holes form a co-dimension 2 ``surface" within the space of general black holes parameterized by $(m,\,a,\,q)$. On any point along the resulting BPS line, the quantum numbers $J^*$ and $\ell_7 Q^*$ are dependent variables: they are both expressed in terms of a single dimensionless parameter which we take as $ag$. 
For this reason, the expression of black hole entropy as a function of conserved charges  \eqref{eq:BHentropy} is not unique. For example, the translation of the parameter $ag$ to the physical variables $J^*$ and $\ell_7 Q^*$ may equally well yield the alternate form
\begin{align}
S^* &=2\pi\sqrt{  N^3 J^*+3 (\ell_7 Q^* )^2 -
	\sqrt{
		\left( N^3 J^*+3 (\ell_7 Q^* )^2\right)^2
		- \frac{2}{3} N^3 (J^*)^3- (\ell_7 Q^*)^4 }
}\,.
\label{eq:BHentropy2}
\end{align}
In the statistical ensemble specified by the quantum numbers $J^*$ and $\ell_7 Q^*$ it is convenient to characterize the BPS line as the vanishing locus of
the ``height" function
\begin{align}
\label{eq:BPS surface}
h & =  \bigg(\left(6  \ell_7 Q-N^3 \right)^2\left((\ell_7 Q)^4 + \frac{2}{3} N^3 J^3\right) +\left[6 (\ell_7 Q^*)^3- 3 N^3 J^2\right]^2 \nonumber\\
&\hspace{2cm} -6\left(6  \ell_7 Q- N^3 \right)\left[ 3 ( \ell_7 Q^*)^2 + N^3 J\right]\left( 2 (\ell_7 Q)^3- N^3 J^2 \right)\bigg)\frac{1}{3 N^3 J^2-6 (\ell_7 Q^*)^3}\,.
\end{align}
For example, non-rotating black holes obviously have angular momentum $J=0$. These are the AdS-Reissner-Nordstr\"{o}m geometries. They include an extremal black hole $T=0$, the lightest in the family that is regular. However, a generic extremal AdS-Reissner-Nordstr\"{o}m black hole is not supersymmetric because $h\neq 0$ for $J=0$.

We conclude this subsection by noting an important corrolary of supersymmetry imposing two conditions: it can be {\it broken} in two independent and complementary ways:
\begin{itemize}
	\item
	{\it Near-extremal} BPS black holes have non-vanishing temperature but they satisfy the constraint $h=0$. 
	\item
	{\it NearBPS} extremal black holes have vanishing temperature but they are not supersymmetric because their charges violate the constraint. 
\end{itemize}
In the following three subsections we first study each of these cases separately and then examine their interplay. 

%%%%%%%%%%%%%%%%%%%%%
\subsection{Near-Extremal Thermodynamics}
In this subsection we consider black holes that depart from BPS by having elevated temperature. This perturbation necessarily increases the mass $M$ from its BPS value $M^*(J^*,Q^*)$. 
However, it does not modify the conserved charges from their reference values $Q^*$ and $J^*$ so they will still be related by the constraint \eqref{eq:BPS surface} that is satisfied on the BPS-line. 

The specific heat $C_T=\frac{dQ}{dT}=T\frac{dS}{dT}$ is the response coefficient that characterizes the increased temperature. At leading order away from extremality the specific heat is linear in temperature so the  derivative $\partial_T S = \frac{C_T}{T}$ (taken with conserved charges held fixed) is a constant.

We have so far analyzed physical variables as functions of the parameters $(m, q, a)$ but the explicit expressions for the entropy $S$ and the potentials $T$, $\Phi$, $\Omega$ \eqref{eq:ST-7} involve the horizon location $r_+$, defined as the largest solution to the horizon equation $Y(r)=0$. This amounts to an unappealing quartic equation that must be solved for $r^2_+$ so it is preferable to eliminate $m$, which appears linearly, in favor of $r_+$. This yields expressions for $S$ , $T$, $\Phi$, and $\Omega$ in terms of parameters $r_+$, $q$ and $a$,
\begin{align}
\label{eqn:Sm}
S&=\frac{\pi ^3 \left(\rho_+^6+2 q \left(r_+^2+a^3 g\right)\right)}{4 G_7 r_+ \Xi^3}\,,\\
\label{eqn:Tm}
T&=\frac{3 g^2 \rho_+^8+ \left(2-6 a^2 g^2\right)\rho_+^6+\left(4 g^2 q-3a^2\Xi\right)\rho_+^4-8 a^2 g^2 q\rho_+^2 -4 a^3 g q\Xi_--4 g^2 q^2}{2\pi r_+ \left(\rho_+^6+2 q \left(r_+^2+a^3 g\right)\right)}\,,\\
\label{eqn:Phm}
\Phi &=\frac{2r_+ \sqrt{2q\left(\rho_+^6 \left(1+g^2 r^2_+\right)+2 q \left(2 g^2 \rho^4+\Xi\rho^2_+ - a^2 \Xi_-^2\right)+4 g^2 q^2\right)}}{\rho_+^6+2 q (r_+^2 + a^3 g )}\,,\\
\label{eqn:Om}
\Omega &=\frac{a \rho_+^4 \left(1+g^2 r_+^2\right)+2 a g q \left(a+g r_+^2\right)}{g \rho_+^6+2 g q \left(r_+^2+a^3 g\right)}\,.
\end{align}
From this starting point it is straightforward to compute the inverse Jacobian 
\begin{align}
\left(\frac{\partial\left(T, Q, J\right)}{\partial\left(a,q,r_+\right)}\right)^{-1}&=\frac{\partial\left(a,q,r_+\right)}{\partial\left(T, Q, J\right)}~,
\end{align}
as well as the derivatives of $S(a,q,r_+)$. Combining these ingredients, we find the specific heat
\begin{align}
\label{eq:CT}
\frac{C_T}{T}&=\left(\frac{\partial S}{\partial a}\right)_{q,r_+}\left(\frac{\partial a}{\partial T}\right)_{Q,J}+\left(\frac{\partial S}{\partial q}\right)_{a,r_+}\left(\frac{\partial q}{\partial T}\right)_{Q,J}+\left(\frac{\partial S}{\partial r_+}\right)_{q,a}\left(\frac{\partial r_+}{\partial T}\right)_{Q,J}\nonumber\\
&= N^3 \frac{32\pi^2 a^4 g^{3}(1+ 10 ag -3 a^2 g^2)}{3(1-a g)^3 (1+3 a g) (3+ 10 ag +19 a^2 g^2)}\,.
\end{align}
This result is positive in the physical regime $0<ag<1$, as required for a stable system. 

As a consistency check, we can instead derive the specific heat by 
considering (\ref{eqn:MnMstar}) for the mass $M$ above its BPS value $M^*$. 
Inserting the parametric form of the mass excess at the second order (\ref{eqn:quadmq}) and the temperature $T$ (\ref{eqn:linearT}), the expression becomes: 
\begin{align}
\label{eq:mvarT}
M-M^* &= \frac{C_T}{2T}\,T^2\,.
\end{align}
This procedure again gives the expression (\ref{eq:CT}) for $\frac{C_T}{T}$, as demanded by the first law of thermodynamics applied with charges kept fixed. 

Yet another way to calculate the specific heat is via the nAttractor mechanism \cite{Larsen:2018iou,Hong:2018viz}. In this simple and illuminating construction
the elevated temperature is taken into account geometrically through the outward displacement of the horizon, without ever deforming away from the BPS geometry. This method yields the more concise expression,
\begin{align}
\frac{C_T}{T}&=\left(\frac{\partial S}{\partial r_+}\right)_{q,a}\left(\frac{\partial T}{\partial r_+}\right)^{-1}_{q,a}~,
\end{align}
which again evaluates to the result in \eqref{eq:CT}. 

We can exploit the nAttractor mechanism taking temperature into account via a simple radial derivative also when analyzing other physical variables. By first 
specializing the potentials $\Phi$, $\Omega$ given in \eqref{eq:ST-7} to the BPS geometry, and only then taking the radial derivative, we easily calculate 
\begin{align}
\label{eq:dtphi}
\left(\frac{\partial \Phi}{\partial T}\right)_{Q,J}&=
\left(\frac{\partial \Phi}{\partial r_+}\right)_{q,a}\left(\frac{\partial T}{\partial r_+}\right)^{-1}_{q,a} = - \frac{12 \pi  a  (1-a g)}{ 3+ 10 ag +19 a^2  g^2}\sqrt{\frac{1+3 a g}{3+a g}}\, , \\
\label{eq:dtomega}
\left(\frac{\partial \Omega}{\partial T}\right)_{Q,J}&=   \left(\frac{\partial \Omega}{\partial r_+}\right)_{q,a}\left(\frac{\partial T}{\partial r_+}\right)^{-1}_{q,a}= - \frac{8 \pi  a g  (1-a g)}{3+10 a g +19 a^2  g^2}\sqrt{\frac{1+3 a g}{3+a g}}\,.
\end{align}
Both of these quantities are negative. It is also noteworthy that they are nearly identical: 
\begin{align}
\label{eq:Tderrel}
2 g \left(\frac{\partial \Phi}{\partial T}\right)_{Q,J}=3\left(\frac{\partial \Omega}{\partial T}\right)_{Q,J}\,.
\end{align} 
The physical interpretation of this relation will be discussed in Section \ref{sec:BPSmic}.

%%%%%%%%%%%%%%%%%%%%%%%%%%%%%
\subsection{Extremal Near-BPS Thermodynamics}
In this subsection, we perturb a BPS configuration by changing charges $J$ and $Q$ so the constraint \eqref{eq:BPS surface} is no longer satisfied but extremality is maintained. Thus the temperature remains zero: the mass is at its minimum possible value, albeit for the ``new" charges. This perturbation is complementary to the temperature/mass deformation studied in the previous subsection. 

The starting point is a BPS state characterized by rotation parameter $a$ and the BPS assignments $\rho_+^{*2}$ \eqref{eq:BPSradius} 
and $q^*$ \eqref{eqn:qstar} corresponding to that value of $a$. Variations $\delta r_+^2 = r_+^2-r_+^{*2}$ and $\delta q = q-q^*$ away from the BPS values generally modify the temperature $T$ to the value given in 
(\ref{eqn:linearT}). It remains zero at linear order exactly when these perturbations are correlated as
\begin{align}
\label{eq:T=0variation}
\delta r^2_+=-\delta q\,\,\frac{g (1+3 a g) (3+ 6 a g +7 a^2g^2)}{2 a (1 + ag)^2 (3+10 ag +19 a^2g^2)}\,.
\end{align}
Furthermore, for variations correlated in precisely this manner (\ref{eqn:quadmq}) yields a greatly simplified formula for the parametric mass
\begin{align}
\label{eqn:mqrela}
m = 3 ag (2 + 3 ag) q
+ \frac{ g^2 (1+3 a g)^4}{2 a^2 (1+ag)^2 (3+ 10 a g  + 19 a^2g^2)}\delta q^2\,,
\end{align}
which expresses the amount that the energy exceeds the BPS mass. The reason for this excess is that the BPS bound cannot be saturated when the constraint is not satisfied.  

A convenient measure of the distance from the BPS line along the
extremal surface is the combination of potentials
\begin{align}
\label{eq:varphidef}
\varphi&= 2(\Phi-\Phi^*) + 3\ell_7(\Omega^*-\Omega)  \,,
\end{align}
which manifestly vanishes on the BPS line where $\Phi=\Phi^*=2$ and $\Omega=\Omega^*=g = \ell_7^{-1}$. 
Using the formulae \eqref{eqn:Phm} and \eqref{eqn:Om} for $\Phi$ and $\Omega$ we expand 
to linear order in $\delta r^2_+$ and $\delta q$, followed by simplification using the relation \eqref{eq:T=0variation}.
This yields the simple expression: 
\begin{align}
\label{eqn:varphidelq}
\varphi&= 2 \frac{\delta q}{q^*}\,.
\end{align}
Thus the composite potential $\varphi$ measures
the relative change of $q$ as it departs from $q^*$ along the extremal surface. Moreover, the physical parameters $M, J, Q$ given in \eqref{eqn:MJQ} are all proportional to $q$ at linear order, 
since \eqref{eqn:mqrela} equates $m$ and $q$ up to terms of quadratic order. We therefore interpret the
potential $\varphi$ as the generator of a scale transformation that acts on the entire black hole geometry, as implemented through rescaling of $M, J, Q$ by a common factor. The numerical factor $2$ in \eqref{eqn:varphidelq} shows that $\varphi$ is $2$ times the relative rescaling of these physical parameters. 

The change of the parameter $q$ as we depart from the BPS line while maintaining zero temperature inevitably changes both the electric potential $\Phi$ and the rotational velocity $\Omega$. We express this dependence through the derivatives 
\begin{align}
\label{eq:dphiphi}
\partial_\varphi\Phi&=\frac{2 a g (1+7 a g)}{3+10 a g + 19 a^2g^2}\,,\\
\label{eq:dphiomega}
\partial_\varphi\Omega &=-\frac{(1-a g) (1+3 a g)g}{3+10 a g +19 a^2g^2}\,.
\end{align}
These relations satisfy
\begin{align}
\label{eq:phiderrel}
2\,\partial_\varphi\Phi -3\,\ell_7\,\partial_\varphi \Omega  =1\,,
\end{align}
as expected from \eqref{eq:varphidef}. Scale transformations with $\varphi>0$ are preferred because they decrease the rotational velocity below $\Omega^*\ell_7=1$ which corresponds to the speed of light in the dual boundary theory. They increase $\Phi$ above its critical value $\Phi^*=2$.

As we have already stressed, the motion away from the BPS line with temperature fixed necessarily increases the energy. In analogy with electrodynamics, the capacitance is the response coefficient measuring energy as $\varphi$ increases. We introduce a coefficient $C_\varphi$ that is proportional to the temperature $T$ (but evaluated as $T\to 0$) through the mass formula
\begin{align}
\label{eq:mvarphi}
M-M^* &= \frac{C_\varphi}{2T}\,\frac{\varphi^2}{(2\pi)^2}\,,
\end{align}
The expression \eqref{eqn:mqrela} for the parametric mass $m$ gives 
\begin{align}
\frac{C_\varphi}T&=N^3 \frac{32\pi^2 a^4 g^{5}(1+ 10 ag -3 a^2 g^2)}{3(1-a g)^3 (1+3 a g) (3+ 10 ag +19 a^2 g^2)}\,.
\end{align}
Alternatively, the changes in potentials \eqref{eq:dphiomega} and the scaling transformations $dQ=\frac12Qd\varphi$, $dJ=\frac12Jd\varphi$ give
\begin{align}
-\frac{C_\varphi}{T}\,\frac{\varphi\, d\varphi}{(2\pi)^2}&=-2(\Phi-\Phi^*)dQ-3(\Omega-\Omega^*)dJ~,
\end{align}
with the expression for the capacitance $C_\varphi$ the same as before, as demanded by
the first law of thermodynamics \eqref{BPS1stlaw}.

We have been careful to define the coefficient $C_\varphi$ entirely through properties within the extremal surface $T=0$. It is therefore interesting 
that numerically
\begin{align}
C_\varphi&= \ell_7^{-2}\, C_T\,,
\end{align}
where $C_T$ is as given in \eqref{eq:CT} and the factors of $\ell_7$ are introduced to account for the different mass dimensions of $[T]=L^{-1}$ and $[\varphi]=L^0$. This relation is expected from $\mathcal{N}=2$ supersymmetry \cite{Fu:2016vas}.

%%%%%%%%%%%%%%%%%%%%%%%%%%%%%%%%%%%%

\subsection{Near-BPS Thermodynamics}
Having explored the two independent deformations of a BPS configuration, in this subsection we put them together to explore the entire near BPS region of parameter space. 

Taking advantage of \eqref{eqn:quadmq}
we expand the mass $M$ around its BPS value and find
\begin{align}
\label{eqn:mmmstar}
M-M^* &=\frac{C_T}{2T}\left[\,T^2 +\frac{\varphi^2}{(2\pi\ell_7)^2}\right]\,.
\end{align}
This is simply a sum of the independent contributions from $T$ \eqref{eq:mvarT} and $\varphi$ \eqref{eq:mvarphi}, with no interplay between the two deformations. We want to understand why the increase in mass takes this form.

We begin by introducing another thermodynamic coefficient, $C_E$. Consider the amount by which the black hole entropy $S$  exceeds its BPS value $S^*$.
At linear order, the difference between the general form of the entropy $S(q,r_+,a)$ \eqref{eqn:Sm}
and its BPS limit $S^*(a)$
\eqref{eq:sbps}
yields terms proportional to $q-q^*$ and $r_+-r^*$. These perturbations are equivalent to the small physical potentials 
$T$ and $\varphi$
via 
 \eqref{eqn:linearT} and \eqref{eqn:varphidelq}. Therefore, the differential change in entropy can be expanded as 
\begin{align}
\label{eq:SBPSvar}
d(S-S^*) &= \frac{C_T}{T}dT + \frac{C_E}{T} \frac{d\varphi}{2\pi}\,.
\end{align}
Explicit computation shows that the coefficient $C_T$ introduced here agrees with its namesake in \eqref{eqn:mmmstar}, as demanded by 
the first law of thermodynamics.
The coefficient of $\varphi$ is a new response coefficient that takes the value
\begin{equation}
    \label{eqn:CE7ref}
\frac{C_E}{T} = \frac{128 \pi^2  a^4 g^4 N^3 (1+a g)^3}{(1-a g)^3 \sqrt{3+a g} (1+3 a g)^{3/2} \left(3+10 a g+19 a^2 g^2\right)}\,.
\end{equation}

Now, we must be careful because $C_E$ is subject to an ambiguity. In the preceding paragraph we  specified for definiteness $S^*$ as the function of $a$ given in
\eqref{eq:sbps}.
Its differential $dS^*$ is proportional to  $da$ which is along the BPS surface. 
However, it may be more appropriate to specify BPS entropy $S^*$ as a function of charges $J, Q$.
The resulting differentials $dJ$, $dQ$ do not generally respect the constraint between charges $h=0$. Therefore, they may include a contribution  normal to the BPS surface, in the direction of   $d\varphi$. Thus the value of $C_E$ depends on the reference point $S^*$.

There is a simple method to take this dependence into account. In the nearBPS regime the parameters $m$ and $q$ are proportional up to quadratic corrections \eqref{eqn:quadmq}. Therefore, in this regime $J$ and $Q$ \eqref{eqn:MJQ} are both functions of $a$, except for an overall factor of $q$. We already noted that the differential $da$ is entirely within the BPS surface. However, $q$ and $\varphi$ are closely related \eqref{eqn:varphidelq} so the overall factor $q$ yields simple additional terms 
\begin{equation}
\label{eqn:dQdJ}
dQ = \frac{1}{2}Q^* d\varphi + \ldots ~~~,~dJ = \frac{1}{2}J^* d\varphi+ \ldots~,
\end{equation}
where the dots refer to the differential $da$ within the BPS surface.
These formulae determine the dependence of $dS^*$ on $d\varphi$ when $S^*$ is presented as a function of the charges $Q$, $J$ rather than just $a$.

The simplest formula for the BPS entropy we know is $S^*(J,Q)$ given in \eqref{eq:BHentropy}. The algorithm in the preceding paragraph easily computes the part of the differential $dS^*$ that is not within the BPS surface. This contribution changes $d(S-S^*)$ by a term that shifts the initial result for $C_E$ given in \eqref{eqn:CE7ref}. It 
becomes  
\begin{align}
\label{eq:CE}
C_E&=N^3\frac{32 \pi ^2 a^4 g^4 \left(1+10 a g-3 a^2 g^2\right) \left(3+21 a g+54 a^2 g^2+66 a^3 g^3-9 a^4 g^4-7 a^5 g^5\right)}{(1-a g)^3 \sqrt{3+a g} (1+3 a g)^{3/2} \left(3+ 10 a g+19 a^2 g^2\right) \left(1-6 a^2 g^2-24 a^3 g^3-3a^4 g^4\right)}\,.
%&=\frac{2\pi}{S^*}\left(\frac{C_T}{T}\right)\frac{\pi ^2 a^4 \left(7 a^5 g^5+9 a^4 g^4-66 a^3 g^3-54 a^2 g^2-21 a g-3\right)}{(1-a g)^3 (3 a g+1)^2 \left(3 a^4 g^4+24 a^3 g^3+6 a^2 g^2-1\right)}.
\end{align}

As we have stressed, the charges 
$Q^*$ and $J^*$ are not independent, they satisfy the constraint $h=0$. 
Therefore, there is no unique way to define the BPS entropy $S^*(J,Q)$ and the response coefficient $C_E$ is sensitive to this ambiguity. For example, for $S^*(J,Q)$ given in \eqref{eq:BHentropy2} we find
\begin{align}
C_E=&\frac{N^3}3\pi ^2\bigg( 27+387 a g+2133 a^2 g^2+5877 a^3 g^3+6798 a^4 g^4-6338 a^5 g^5-21302 a^6 g^6\nonumber\\
&-5910 a^7 g^7+48615 a^8 g^8+7695 a^9 g^9-4527 a^{10} g^{10}-687 a^{11} g^{11}\bigg)\left(3+ 10 a g+19 a^2 g^2\right)^{-1}\nonumber\\
&\times\left((1-a g)^3
   \sqrt{(1+ 3 a g)(3 +ag)}\left(3+21 a g+54 a^2 g^2+66 a^3 g^3-9 a^4 g^4-7 a^5 g^5\right)\right)^{-1}\,.
\end{align}
This coefficient is always positive in the entire regime $0<ag<1$. This designation of reference $S^*$ assigns positive entropy to all perturbations with $\varphi\ge 0$.

The method for analyzing $dS^*$ can be applied to other functions of $(J, Q)$ as well. An important example is the height function $h$ itself. Any surface $h={\rm constant}$ 
is characterized by the one-form $dh$ and the value of the ``constant" measures its distance from the BPS surface $h=0$. The potential $\varphi$ is another such measure so, according to the general theory of surfaces, $h$ must be proportional to $\varphi$ at linear order. The constant of proportionality follows from the rule 
\eqref{eqn:dQdJ}:
\begin{align}
\label{eq:adef2}
h=& \frac{1}{9(J^2 N^3 - 2 (Q\ell_7)^3)}\bigg(27 J^6 N^9+4 J^5 N^6 \left(-2 N^6+21 N^3 \ell_7Q+27 (\ell_7Q)^2\right)\nonumber\\
&+27 J^4 N^6(\ell_7Q)^2 \left(11 \ell_7Q-2 N^3\right)+12 J^3 N^3 (\ell_7Q)^3 \left(3 N^6-34 N^3 \ell_7Q-12 (\ell_7Q)^2\right)\nonumber\\
&+3 J^2 N^3 (\ell_7Q)^4 \left(N^6+54 N^3 \ell_7Q-432 (\ell_7Q)^2\right)-36 J N^3 (\ell_7Q)^6 \left(N^3-12 \ell_7Q\right)\nonumber\\
&-3 (\ell_7Q)^7 \left(N^6+48 N^3 \ell_7Q-
432(\ell_7Q)^2\right)\bigg)\varphi
\\
\label{eq:adef}
=&\frac{128 a^4 g^4 N^9  (1+a g)^8 \left(1+10 a g -3 a^2 g^2\right)}{9 (1-a g)^9 (3+a g) (1+3 a g)^4}\varphi \equiv \alpha\varphi\,,
\end{align}
The constant 
$\alpha$ is strictly positive in the entire regime $0<ag<1$. 

We are now in a position to better understand the first law of thermodynamics in the near BPS region:
\begin{align}
T dS&= d(M-M^*)-2(\Phi-\Phi^*)dQ-3(\Omega-\Omega^*)dJ\,.
\end{align}
The departures of the potentials $\Phi$ and $\Omega$ from their BPS values are given by the sum of their extremal near-BPS (\ref{eq:dphiphi}-\ref{eq:dphiomega}) and BPS near-extremal (\ref{eq:dtphi}-\ref{eq:dtomega}) contributions. Alternatively, $\Phi-\Phi^*$ and $\Omega-\Omega^*$ can be calculated from their general forms (\ref{eqn:Phm}-\ref{eqn:Om}), expanding to linear order in $r_+-r^*$ and $q-q^*$ and then trade these variables for $T$ and $\varphi$. 
We already presented a convenient rule
\eqref{eqn:dQdJ}
that relates the differentials $dQ$ and $dJ$ to the potential $d\varphi$ so we can present the result as
\begin{align}
\label{eq:bpslaw1}
T dS^*+2(\Phi-\Phi^*)dQ+3(\Omega-\Omega^*)dJ&= \left[-\frac{C_E}{T}T + \frac{C_\varphi}{T}~. \frac{\varphi}{2\pi}\right]\frac{d\varphi}{2\pi}\,.
%&=\left[-\frac{C_E}{T}T + \frac{C_\varphi}{T} \frac{\varphi}{(2\pi)}\right]\alpha^{-1}\frac{dh}{2\pi}\,.
\end{align}
The values of the linear response coefficients $C_E$ and $C_\varphi$ that follow from the computation here agree with those given earlier. As we stressed previously, the precise value of $C_E$ depends on the reference value for the BPS entropy $S^*$. We added the term $TdS^*$ so the equality holds no matter the reference value $S^*$, as long as it is consistently applied. 

Combining the partial result above with \eqref{eq:SBPSvar} we find
\begin{align}
TdS&=T d(S-S^*)+TdS^*\nonumber\\
&=\left[\frac{C_T}{T}TdT + \frac{C_E}{T} T\frac{d\varphi}{(2\pi)}\right]+\left[-\frac{C_E}{T}T + \frac{C_\varphi}{T} \frac{\varphi}{(2\pi)}\right]d\varphi-2(\Phi-\Phi^*)dQ\nonumber\\
&\hspace{9cm}-3(\Omega-\Omega^*)dJ~.
\end{align}
The terms with the coefficient $C_E$ cancel precisely and the remaining terms satisfy the first law with the mass term 
$$
M-M^*=\frac{C_T}{2T}\left[T^2 +\frac{\varphi^2}{(2\pi\ell_7)^2}\right]\,.
$$
as expected.

The final physical variable we will study is
Gibbs free energy 
$$
\mathcal{F} = M-2\Phi Q-3\Omega J-TS = M - M^* -2(\Phi-\Phi^*) Q-3(\Omega-\Omega^*)J-TS\,,
$$
particularly its dependence on $T$ and $\varphi$ in the nearBPS limit. 
Since the excitation energy \eqref{eqn:mmmstar} is quadratic in $T$ and $\varphi$,
the first term is negligible. Moreover, since $\Phi-\Phi^*$ and $\Omega-\Omega^*$ are first order in $T$ and $\varphi$, we can replace $Q$, $J$ and $S$ by their BPS values when computing the free energy at first order. The full expression then becomes
\begin{align}
\mathcal{F}=&-2(\Phi-\Phi^*)Q^*-3(\Omega-\Omega^*)J^*-TS^*\\
=&-N^3 \frac{16 a^4  g^5 (1+10 a g -3 a^2 g^2)}{3(1-a g)^3 (1+3 a g) (3+ 10 a g + 19 a^2 g^2)}\, \varphi \nonumber\\
&\hspace{3cm}+N^3\frac{32 \pi a^4 g^4 (3+3 a g-31 a^2 g^2-7 a^3 g^3)}{3 (1-a g)^3 \sqrt{3+a g} (1+3 a g)^{3/2}(3+ 10 a g + 19 a^2 g^2)}\, T\,.
\end{align}
Here the coefficient of $\varphi$ is always negative while the coefficient of $T$ switches sign depending on the value of $a$. 

We will now proceed to study the free energy from a microscopic point of view.

%%%%%%%%%%%%%%%%%%%%%%%%%%%%%%%%%%%%
\subsection{BPS Entropy Extremization}
\label{sec:BPSmic}

 The partition function $Z$ of a black hole is defined in Euclidean quantum gravity as the on-shell action. It depends on potentials that are specified as asymptotic boundary conditions on spacetime. While it is a thermodynamic quantity in gravity, it is identified in the dual microscopic description as a trace over quantum states that, in the context of AdS$_7$, we introduce as 
\begin{align}
\label{eqn:ZAdS7}
Z = {\rm Tr} ~e^{-\beta (M - 2\Phi Q - 3\Omega J)}
= {\rm Tr} ~e^{-\beta [(M-M^*) - 2\beta(\Phi-\Phi^*) Q - 3\beta(\Omega-\Omega^*) J)}~,
\end{align}
where the BPS mass is $M^*=2\Phi^* Q+3\omega^* J$. Here and in all microscopic considerations that follow in the following subsections we simplify units so $M$ and $Q$ are dimensionless. This amounts to taking $\ell_7=g^{-1}=1$. 

In this subsection we want to compare our gravitational results with microscopic ideas in the BPS limit
which involves, in particular, the extremal limit $\beta\to\infty$ where the gravitational potentials $\Phi$, $\Omega$ approach $\Phi^*=2$, $\Omega^*=1$. It is therefore natural to introduce dimensionless extremal potentials
$\Delta = \beta(\Phi-\Phi^*) = \partial_T\Phi$ and $\omega = \beta(\Omega-\Omega^*)=\partial_T\Omega$ that are conjugates to the charges $Q$ and $J$, respectively. 

However, as we have stressed repeatedly, the BPS limit is stronger than extremality. In the microscopic theory
supersymmetry is conveniently implemented as an index that can be defined as 
the complex locus
\begin{equation}
    \label{eqn:indexconstraint}
    2\Delta-3\omega=2\pi i~. 
\end{equation}
The BPS partition function therefore becomes a function of two {\it complex} potentials and only their real parts can be identified with gravitational potentials through ${\rm Re} \Delta=\partial_T\Phi$ and ${\rm Re}\omega=\partial_T\Omega$. 

For the Kerr-Newman AdS$_7$ black hole we study in this section, microscopic considerations give the BPS partition function 
\cite{Nahmgoong:2019hko,Kantor:2019lfo,Benini:2019dyp,Lee:2020rns,Zaffaroni:2019dhb}
\begin{align}
\ln Z (\Delta,\omega) &=\frac{1}{24} N^3 \frac{\Delta^4}{\omega^3}\,.
\end{align}
The black hole entropy is defined in the microcanonical ensemble where conserved charges are specified.
The Legendre transform from the canonical ensemble is conveniently implemented by the entropy function 
\begin{align}
S=\frac{1}{24} N^3  \frac{\Delta^4}{\omega^3} - 2\Delta Q - 3\omega J + \Lambda(2\Delta-3\omega-2\pi i)\,,
\end{align}
where $\Lambda$ is a Lagrange multiplier that enforces the constraint
\eqref{eqn:indexconstraint} on the potential.
The extremization conditions of the entropy function are 
\begin{align}
\label{eq:ext1}
\partial_\Delta S &=\frac{1}{6} N^3 \frac{\Delta^3}{\omega^3}-2 Q + 2\Lambda=0\,,\\
\label{eq:ext2}
\partial_\omega S &= -\frac{1}{8} N^3 \frac{\Delta^4}{\omega^4}-3 J - 3\Lambda=0\,,\\
\label{eq:ext3}
\partial_\Lambda S &=2\Delta-3\omega-2\pi i=0\,.
\end{align}
These equations simplify the entropy function at its extremum so
\begin{equation}
\label{eqn:ads7bhent}
S=-2\pi i \Lambda\,,
\end{equation}
and show that the Lagrange multiplier $\Lambda$ satisfies the quartic equation
\begin{align}
(\Lambda-Q)^4+\frac{2}{3}N^3 (J+\Lambda)^3\label{eq:lambda_eq}
& = \Lambda^4+A\Lambda^3+B\Lambda^2+C\Lambda+D = 0 \,,
\end{align}
with the coefficients
\begin{eqnarray}
\label{eqn:Adef}
A=& \frac{2}{3}N^3-4Q&=\frac{2}{3}N^3 \left(1-\frac{32 a^3 g^3}{(1-ag)^3 (1+3 a g)}\right)\,,\\
\label{eqn:Bdef}
B=& 6 Q^2+2 N^3 J&=\frac{2}{3}N^6 \frac{ (2 a g)^4(1+a g)^4}{(1+3 a g)^2 (1-a g)^6}\,,\\
\label{eqn:Cdef}
C=& 2 N^3 J^2 - 4 Q^3&=\frac{2}{27} N^9 \frac{ (2ag)^8 (3+a g) \left(1-6 a^2 g^2-24 a^3 g^3- 3a^4 g^4\right)}{(1+3 a g)^4 (1-ag)^9}\,,\\
\label{eqn:Ddef}
D=& \frac{2}{3} N^3 J^3 - Q^4&= \frac{2}{81} N^{12} \frac{ (2a g)^{12} \left(3+21 a g+54 a^2 g^2+66 a^3 g^3+15 a^4 g^4+a^5 g^5\right)}{(1+3 a g)^6 (1-a g)^{12}}\,.
\end{eqnarray}
 For each coefficient the second expression introduces the dimensionless parameter $ag$ by rewriting the conserved charges using \eqref{eq:EJQbps}.

All the coefficients in the quartic equation \eqref{eq:lambda_eq} are real so,
for the entropy \eqref{eqn:ads7bhent} to be real, the polynomial must have at least one pair of purely imaginary conjugate roots. Therefore, it must take the form
\begin{align}
\label{eq:lambda_eq2}
\left(\Lambda+\alpha\right)\left(\Lambda+\beta\right)\left(\Lambda^2+\gamma\right)=0\,.
\end{align}
Comparing the coefficients in this quartic polynomial with those of \eqref{eq:lambda_eq} gives $\alpha+\beta=A$ and 
\begin{equation}
    \label{eqn:LambdaAC}
    \Lambda= i \gamma= i \sqrt{\frac{C}{A}}~,
\end{equation}
from the odd powers of $\Lambda$. The even powers similarly yield 
the product $\alpha\beta$ and the consistency condition
\begin{align}
\label{eqn:abcdconst}
A(BC- A D)&=C^2\,.
\end{align}
These identifications determine the BPS entropy. 
The formula for $\gamma$ gives the simplest expression 
and the
consistency condition \eqref{eqn:abcdconst} gives an interesting alternate form: 
\begin{align}
S^* = 2\pi \sqrt{\frac{C}{A}} = {2\pi\sqrt{\frac12\left(B-\sqrt{B^2-4 D}\right)}}~.
\end{align}
The coefficients $A$-$D$ defined in (\ref{eqn:Adef}-\ref{eqn:Ddef}) 
are such that
the two expressions agree precisely with \eqref{eq:BHentropy} and \eqref{eq:BHentropy2}.
Moreover, the consistency condition  \eqref{eqn:abcdconst}
is exactly the constraint on charges \eqref{eq:BPS surface} that is required for supersymmetry. 

The derivation of the intricate formulae for the entropy and the constraint from a simple free energy is very suggestive but not entirely satisfying. The requirement that the entropy be real is essential for reaching these results but there is no clear reason that the Legendre transform cannot be dominated by a complex saddle-point. The reality condition on the entropy is a prescription that is evidently correct but it still awaits a principled physical explanation. 

%%%%%%%%%%%%%%%%%%%%
\subsection{NearBPS Potentials}
\label{sec:ads7nearbps}

The entropy extremization principle for BPS black holes can be leveraged 
to describe nearBPS physics as well. For example, in the following subsection, we will be able to derive the specific heat $C_T$ and the linear response coefficient $C_E$ defined from gravity in \eqref{eq:CECT}. These parameters pertain unambiguously to physical properties away from the BPS surface. 

The extremization of the BPS entropy function in the preceding subsection determines the potentials $\Delta$ and $\omega$ at the extremum: 
\begin{align}
\Delta &=4\pi i\frac{J+\Lambda}{4J+3Q+\Lambda}\,,
\label{eq:delext}
\\
\omega &=-2\pi i\frac{Q-\Lambda}{4J+3Q+\Lambda}\,.
\label{eq:omext}
\end{align}
Here the purely imaginary value of the Lagrange multiplier $\Lambda = i \sqrt{\frac{C}{A}}$ is understood.

The BPS potentials $\Delta$, $\omega$ are genuinely complex, they have nontrivial real and imaginary parts. As we discussed below
\eqref{eqn:indexconstraint} in the extremal (but not necessarily BPS) limit $\beta\to\infty$ we identify their real parts with the thermal derivatives $\partial_T\Phi$ and $\partial_T\Omega$, respectively. A more general departure from the BPS surface that allows temperature as well as violation of the constraint is described by the complex variable $\varphi + 2\pi i T$. This suggests identifying the complex potentials $\Delta$, $\omega$ as derivatives of $\Phi$ and $\Omega$ with respect to $\varphi + 2\pi i T$. This rule yields the 
real parts
\begin{align}
\label{eq:Reparts2}
\text{Re }\Delta&=-\frac{12 \pi  a g (1-a g) }{3+10 a g +19 a^2g^2}\sqrt{\frac{1+3ag}{3+ag}}=\partial_T\Phi\,,\\
\text{Re }\omega&=-\frac{8 \pi  a g (1-a g) }{3+10 a g +19 a^2g^2}\sqrt{\frac{1+3ag}{3+ag}}=\partial_T\Omega\,,
\label{eq:Reparts}
\end{align}
which indeed match the gravitational results (\ref{eq:dtphi}-\ref{eq:dtomega}) exactly.
The imaginary parts of the potentials (\ref{eq:delext}-\ref{eq:omext})
similarly permit the complementary identification 
\begin{align}
\label{eq:Imparts2}
\text{Im }\Delta&=\frac{4 \pi  a g (1+7 a g)}{3+ 10 a g +19 (a g)^2}=2\pi\partial_\varphi\Phi\,,\\
\text{Im }\omega&=-\frac{2 \pi  (1+3 a g) (1-a g)}{3+10 a g +19 (a g)^2}=2\pi\partial_\varphi\Omega\,,
\label{eq:Imparts}
\end{align}
which agree precisely with the gravitational expressions (\ref{eq:dphiphi}-\ref{eq:dphiomega}).  

The reasoning leading us to the real and imaginary parts of $\Delta,\omega$ takes advantage of the complex formulae (\ref{eq:delext}-\ref{eq:omext}) which are cast in terms of conserved charges $Q$ and $J$. We have refrained from employing the charges $Q, J$ in those formulae, opting instead for the coordinate $ag$ 
on the BPS locus that facilitates comparisons with gravity. 
Conceptually, it may seem preferable to retain conserved charges in all final results. This inclination is equivalent to finding a function $a(J,Q)$, with the understanding that the result is 
non-unique due to the BPS constraint $h=0$. It is not difficult to do so, at least in principle. 
For example, we can rearrange \eqref{eq:Imparts2} as a quadratic equation for $ag$ with the solution 
\begin{align}
ag&=\frac{2 \pi-5 \left(\text{Im}\Delta\right) - 2 \sqrt{\pi ^2+16 \pi  \left(\text{Im}\Delta\right) -8 \left(\text{Im}\Delta\right)^2}}{28 \pi  g-19 \left(\text{Im}\Delta\right)g}~,
\end{align}
where, from \eqref{eq:delext}, we have
$$
{\rm Im}\,\Delta = 4\pi \frac{J(4J+3Q)-\Lambda^2}{(4J+3Q)^2-\Lambda^2} ~.
$$
While this is a closed form for $a$ which can in principle give us all BPS potentials and conserved charges in terms of $J$ and $Q$, the expressions are messy and do not seem illuminating. We will arrive at a more concise expression below using the first law of thermodynamics.

%%%%%%%%%%%%%%%%
\subsection{NearBPS Entropy}
\label{subsec:mic7}
In this subsection we introduce a near AdS extremization principle that will account for the entropy in the near BPS region. We follow the AdS$_4$  discussion in subsection \ref{sec:ads4nearbps}. Thus we take the configuration space 
identified by BPS considerations for granted. However, noting that the physical BPS states result from a larger phase space upon imposing a constraint, we consider the additional states that result by  relaxing the constraint from its strict BPS version \eqref{eqn:indexconstraint} to
\begin{align}
\label{eq:ads7bc}
    2(\Phi-\Phi^*)-3(\Omega-\Omega^*)=\varphi+2\pi iT \,.
\end{align}
We implement this by extremizing the  nearBPS free energy,
\begin{align}
T S(\Phi_I,\Omega,\Lambda) = \ln Z = 
& \frac{1}{24} N^3  \frac{(\Phi-\Phi^*)^4}{(\Omega-\Omega^*)^3} - 2(\Phi-\Phi^*) Q - 3(\Omega-\Omega^*) J\nonumber\\
&+ \Lambda(2(\Phi-\Phi^*)-3(\Omega-\Omega^*)-\varphi-2\pi iT )\,. 
\label{eq:nBPSentFct}
\end{align}
The primary change from the BPS case is that now it is the boundary conditions
\eqref{eq:ads7bc} that are imposed by the Lagrange multiplier $\Lambda$. Accordingly, the extremization equations from the BPS considerations
(\ref{eq:ext1}-\ref{eq:ext2}) are unchanged, except for the constraint\eqref{eq:ext3} that is modified to \eqref{eq:ads7bc}.
The extremal value of the free energy is then
\begin{align}
T S=-(\varphi + 2\pi i T)\Lambda\,. 
\end{align}

The significant difference between the BPS and nearBPS cases is that the latter  does not guarantee a purely imaginary root of the quartic equation \eqref{eq:lambda_eq} so the factorized form \eqref{eq:lambda_eq2} does not apply in general. 
Without making any assumptions on coefficients, 
we can rewrite the quartic equation \eqref{eq:lambda_eq}
as
\begin{align}
    \left(\Lambda ^2+\frac{C}{A}\right) \left((\Lambda+\frac{A}{2})^2-\frac{A^3-4 A B+4 C}{4A} \right) &=-\frac{2}{3}\frac{C\, h}{A^2}\,,
\end{align}
where the height function 
\begin{align}
    h=\frac{2}{3}\frac{A(AD-BC)+C^2}{C}\,,
\end{align}
is generally  non-vanishing. Upon inserting expressions for the coefficients $A, B, C, D$ given in (\ref{eqn:Adef}-\ref{eqn:Ddef}), it can be verified that the height function here agrees with the one previously introduced \eqref{eq:BPS surface}. 

In this form of the quartic equation it is manifest that when the 
constraint $h=0$ is satisfied the quartic has a purely imaginary root which agree with the one we already found from the entropy extremization in the BPS case \eqref{eqn:LambdaAC}.
For small violations of the constraint $h=0$, we can perturb around the BPS solution 
and find the shift in the position of the root
\begin{align}
    \delta\Lambda&=\frac{C\,h}{3 A \left(i(A B-2C) \sqrt{\frac{C}{A}}+A C\right)}
    ~.
\end{align}
In the extremal case the height function $h$ is related to the symmetry breaking parameter as
$h=\alpha\varphi$, where the proportionality constant $\alpha$
%\begin{align}
%\alpha=&2 \left(9 J^4 N^6+J^3 \left(-4 N^9+28 N^6 Q+30 N^3  Q\right)-18 J^2 N^3 Q^2 %\left(N^3-5  Q\right)\right.\nonumber\\
%&\left.+6 J N^3 Q^3 \left(2 N^3-15 Q\right)+Q^4
%   \left(N^3-6 Q\right) \left(N^3+36 Q\right)\right)\,.
%\end{align}
was previously defined in \eqref{eq:adef2}.  
Nonvanishing temperature can be taken into account by the complexification $\varphi\to\varphi + 2\pi i T$ so
the black hole entropy above and beyond the BPS contribution becomes
\begin{align}
    S-S^*=& {\rm Re} \left[ \frac{2\pi i\delta \Lambda \alpha}{h}  (\varphi + 2\pi i T)\right]  \\
    =&\frac{3 S^*(J,Q)\left(3 J^2 N^3-J N^6+6 J N^3 Q-3 N^3 Q^2+12 Q^3\right)\alpha}{4d}\ \varphi\nonumber\\
    &-\frac{3 \pi ^2 \left(N^3-6 Q\right) \left(J^2 N^3-2 Q^3\right) \alpha}{d}\ T\\
    =&\frac{C_E}{T}\frac{\varphi}{2\pi} + \frac{C_T}{T}\,T\,,
    \label{eq:CECT}
\end{align}
where the BPS entropy $S^*$ used for reference is \eqref{eq:BHentropy} and we introduced the notation
\begin{align}
d=&27 J^4 N^6-18 J^3 \left(N^9-6 N^6 Q\right)+2 J^2 N^6 \left(N^3-9 Q\right) \left(2 N^3-9 Q\right)\nonumber\\
&+18 J N^3 Q^2 \left(N^3-6 Q\right) \left(N^3-4 Q\right)+Q^3 \left(-2 N^9+63 N^6 Q-432 N^3 Q^2+864Q^3\right)\,.
\end{align}
The excess entropy $S-S^*$ we find here takes the same form as the
gravitational formula \eqref{eq:SBPSvar}. Moreover, the linear response coefficients $\frac{C_E}{T}$ \eqref{eq:CE} and  $\frac{C_T}{T}$ \eqref{eq:CT} agree exactly. This match is between rather elaborate functions and involves physical variables that break supersymmetry. 

The extremization conditions on the potentials are the same as their BPS analogues (\ref{eq:ext1}-\ref{eq:ext2}), after simple substitutions in the notation. Therefore, the solutions can be adapted 
from (\ref{eq:delext}-\ref{eq:omext}) with minimal effort. The key step is that we now impose the constraint
\eqref{eq:ads7bc} and then identify the physical potential as the real part of the result. For the rotational velocity we find  
\begin{align}
    \text{Re}\,\left(\Omega-\Omega^*\right)& = - \text{Re }\left[\frac{Q-\Lambda}{4J + 3Q + \Lambda}(\varphi+2\pi i T)\right] \nonumber\\
%    &=\frac{ -Q(4 J + 3 Q) - \Lambda^2}{(4 J+3 Q)^2-\Lambda ^2} \varphi +2\pi i\frac{\Lambda (4 J + 4 Q)}{(4 J+3 Q)^2-\Lambda ^2}T\nonumber\\
    &=-\frac{(1-ag) (1+3 a g)}{3+10ag+19 a^2 g^2}\varphi-\frac{8\pi a g (1-a g)}{\left(3+10 a g+19 a^2 g^2\right)}\frac{\sqrt{1+3 a g}}{\sqrt{3+a g}}T~,
    %\nonumber\\
%    &=\frac{\text{Im}\,\omega}{2\pi}\varphi-\text{Re}\,\omega\, T=\partial_\varphi(\text{Re }\Omega)\,\varphi-\partial_T(\text{Re }\Omega)\,T\,.
\end{align}
and the electric potential similarly
yields
\begin{align}
    \text{Re}\,\left(\Phi-\Phi^*\right)&=
    %\frac{\text{Im}\,\Delta}{2\pi}\varphi-\text{Re}\,\Delta\, T=\partial_\varphi(\text{Re }\Phi)\,\varphi-\partial_T(\text{Re }\Phi)\,T\,.
    \text{Re }\left[\frac{2(J+\Lambda)}{4J + 3Q + \Lambda}(\varphi+2\pi i T)\right]\\
    &=\frac{2 a g (1+7 a g)}{3+10 a g + 19 a^2g^2}\varphi-\frac{12 \pi  a  (1-a g)}{ 3+ 10 ag +19 a^2  g^2}\sqrt{\frac{1+3 a g}{3+a g}}T~.
\end{align}
The thermal dependences in these formulae agree with the gravitational potentials 
(\ref{eq:dtphi}-\ref{eq:dtomega}) and the 
constraint violations similarly reproduce
(\ref{eq:dphiphi}-\ref{eq:dphiomega}).
Thus the near BPS extremization principle gives the correct physical potentials. Of course these agreements are also closely related to the analogous results for BPS potentials (\ref{eq:Reparts2}-\ref{eq:Imparts}).
 
Let us finally comment that the boundary conditions
\eqref{eq:ads7bc} on the complex potentials give  
\begin{align}
2\text{Re}\,\left(\Phi-\Phi^*\right)-3\text{Re}\,\left(\Omega-\Omega^*\right)=\varphi~.
\end{align}
Since we identify the gravitational potentials with these real parts we immediately find the simple relations
\begin{align}
%    2\text{Im}\,\Delta-3\text{Im}\,\omega=2\pi &\Rightarrow 
2\partial_\varphi\Phi&=3\partial_\varphi\Omega + 1 \,,\\
%    2\text{Re}\,\Delta-3\text{Re}\,\omega=0 &\Rightarrow 
2\partial_T\Phi &=3\partial_T\Omega \,.
\end{align}
These equations exactly match \eqref{eq:Tderrel}
and \eqref{eq:phiderrel} that we found in the gravitational calculation.

\section*{Acknowledgements}
%We thank NN for useful discussions. 
FL thanks Sangmin Choi, Nizar Ezroura, Siuyl Lee, Zhihan Liu, Jun Nian, and Yangwenxiao Zeng for collaboration on related research. 

This work was supported in part by the U.S. Department of Energy under grant DE-SC0007859. The work of SP was supported in part by a Barbour Fellowship from the University of Michigan.

%%%%%%%%%%%%%%%%%%%%%%%%%%%%%%
%%%%%%%%%%%%%%%%%%%%%%%%%%%%%%
\bibliographystyle{JHEP}
\bibliography{AdSBib.bib}

%%%%%%%%%%%%%%%%%%%%%%%%%%%%%%
%%%%%%%%%%%%%%%%%%%%%%%%%%%%%%

\end{document}